\documentclass[aps,prx,twocolumn,superscriptaddress,nofootinbib,floatfix,showkeys,longbibliography]{revtex4-2}

\usepackage{graphicx}  
\usepackage{subfigure}
\usepackage{dcolumn} 
\usepackage{bm}
\usepackage{amssymb}   
\usepackage{amsfonts}   
\usepackage{comment}
\usepackage[colorlinks,linkcolor=blue,anchorcolor=blue,citecolor=blue,urlcolor=blue]{hyperref}
\usepackage{xcolor} 
\usepackage{soul}
\usepackage{tabularx,booktabs,ragged2e,multirow}
\newcolumntype{C}{>{\centering\arraybackslash}X}
\usepackage{multirow}
\usepackage{amsmath}
\usepackage{mathrsfs}
\usepackage{mathcomp}
\usepackage{textcomp}
\usepackage{dsfont}
\usepackage{esint}
\usepackage{braket}
\usepackage{textgreek}
\usepackage{lipsum}
\usepackage{marvosym} 
\usepackage{centernot}
\usepackage{cancel}
\usepackage{dashrule}
\usepackage{enumitem}
\usepackage{tikz}
\usepackage[compat=1.1.0]{tikz-feynman}
\def\shrinkage{0mu}
\def\vecsign{\mathchar"017E}
\def\dvecsign{\smash{\stackon[-1.95pt]{\mkern-\shrinkage\vecsign}{\rotatebox{180}{$\mkern-\shrinkage\vecsign$}}}}
\def\dvec#1{\def\useanchorwidth{T}\stackon[-4.2pt]{#1}{\,\dvecsign}}
\usepackage{stackengine}
\stackMath

\DeclareMathOperator{\sgn}{sgn}

\begin{document}

\title{Floquet spin textures in optically pumped non-Hermitian surface states}

\author{Xiao-Xiao Zhang}
\email{xxzhang@hust.edu.cn}
\affiliation{Wuhan National High Magnetic Field Center and School of Physics, Huazhong University of Science and Technology, Wuhan 430074, China}

\author{Naoto Nagaosa}
\affiliation{RIKEN Center for Emergent Matter Science (CEMS), Wako, Saitama 351-0198, Japan}
\affiliation{Fundamental Quantum Science Program, TRIP Headquarters, RIKEN, Wako, Saitama 351-0198, Japan}


\newcommand{\ba}{{\bm a}}
\newcommand{\bd}{{\bm d}}
\newcommand{\bb}{{\bm b}}
\newcommand{\bk}{{\bm k}}
\newcommand{\bmm}{{\bm m}}
\newcommand{\bn}{{\bm n}}
\newcommand{\br}{{\bm r}}
\newcommand{\bq}{{\bm q}}
\newcommand{\bp}{{\bm p}}
\newcommand{\bv}{{\bm v}}
\newcommand{\bA}{{\bm A}}
\newcommand{\bB}{{\bm B}}
\newcommand{\bD}{{\bm D}}
\newcommand{\bE}{{\bm E}}
\newcommand{\bH}{{\bm H}}
\newcommand{\bJ}{{\bm J}}
\newcommand{\bK}{{\bm K}}
\newcommand{\bL}{{\bm L}}
\newcommand{\bP}{{\bm P}}
\newcommand{\bS}{{\bm S}}
\newcommand{\bX}{{\bm X}}
\newcommand{\brho}{{\bm \rho}}
\newcommand{\cA}{{\mathcal A}}
\newcommand{\cB}{{\mathcal B}}
\newcommand{\cG}{{\mathcal G}}
\newcommand{\cM}{{\mathcal M}}
\newcommand{\cP}{{\mathcal P}}
\newcommand{\cT}{{\mathcal T}}
\newcommand{\bdelta}{{\bm \delta}}
\newcommand{\bgamma}{{\bm \gamma}}
\newcommand{\bGamma}{{\bm \Gamma}}
\newcommand{\bzero}{{\bm 0}}
\newcommand{\bOmega}{{\bm \Omega}}
\newcommand{\bsigma}{{\bm \sigma}}
\newcommand{\bUpsilon}{{\bm \Upsilon}}
\newcommand{\bcA}{{\bm {\mathcal A}}}
\newcommand{\bcB}{{\bm {\mathcal B}}}
\newcommand{\bcD}{{\bm {\mathcal D}}}
\newcommand\dd{\mathrm{d}}
\newcommand\ii{\mathrm{i}}
\newcommand\ee{\mathrm{e}}
\newcommand\zz{\mathtt{z}}
\newcommand\cE{\mathcal{E}}
\newcommand\cD{\mathcal{D}}
\newcommand\colonprod{\!:\!}

\makeatletter
\let\newtitle\@title
\let\newauthor\@author
\def\ExtendSymbol#1#2#3#4#5{\ext@arrow 0099{\arrowfill@#1#2#3}{#4}{#5}}
\newcommand\LongEqual[2][]{\ExtendSymbol{=}{=}{=}{#1}{#2}}
\newcommand\LongArrow[2][]{\ExtendSymbol{-}{-}{\rightarrow}{#1}{#2}}
\newcommand{\cev}[1]{\reflectbox{\ensuremath{\vec{\reflectbox{\ensuremath{#1}}}}}}
\newcommand{\red}[1]{{\leavevmode\color{red}#1}} 
\newcommand{\blue}[1]{{\leavevmode\color{blue}#1}} 
\newcommand{\green}[1]{\textcolor{orange}{#1}} 
\newcommand{\mytitle}[1]{\textcolor{orange}{\textit{#1}}}
\newcommand{\mycomment}[1]{} 
\newcommand{\note}[1]{ \textbf{\color{blue}#1}}
\newcommand{\warn}[1]{ \textbf{\color{red}#1}}

\makeatother

\begin{abstract}
Optical effects of quantum matter with interaction are key to physics and technology. The class of non-Hermitian (NH) phenomena is mostly explored in cold atoms, photonics, and metamaterials out of equilibrium. 
Effective NH systems due to interaction in equilibrium solids, however, provide a unique opportunity for realizing light-NH matter hybrids via periodic irradiation. Given NH topological surface states with magnetic disorder, here we reveal spectroscopically observable Floquet spin textures for this hybrid quantum matter: merged meron strings, dichroic skyrmions, and domain structure accompanying vortices in energy planes; Bloch lines and topologically twisted vortex rings distinct from Hopfions in energy-momentum space; topological selectiveness of linear polarization occurs besides circular dichroism.
Spectroscopic energy evolution and chemical potential dependence are key ingredients of this open hybrid system. With spin-resolved photoemission spectroscopy, this scenario bears prime physical interest by reaching the crossroad between solid-state interaction effects, non-Hermiticity, and light-matter coupling.
\end{abstract}

\maketitle

\let\oldaddcontentsline\addcontentsline
\renewcommand{\addcontentsline}[3]{}

\section*{Introduction}\label{Sec:intro}
Light-matter coupled systems are a constant source of intriguing phenomena and advancements for information application, especially in the rapidly growing paradigm where controlled light-matter interaction promises to synthesize unconventional quantum states\cite{Giannetti2016,Nicoletti2016,Bloch2022}. For instance, periodic pumping of equilibrium conventional systems leads to the Floquet engineering of topology and symmetry, which has been proved fruitful both theoretically and experimentally in solid-state and broader metamaterial platforms\cite{Oka2009,FloquetTI,Gedik,Gedik3,Aeschlimann2021,Zhou2023,Ito2023,Oka2019,Harper2020,Rudner2020,Bao2021,Torre2021,Mori2023}.
On the map of light-matter coupling or correlated electron-photon systems, this sits, in terms of photon number and coupling strength as two coordinates, at large photon number and relatively weak coupling of conventional matter to light\cite{Oka2019,Bloch2022}. 
Pushing Floquet engineering to a third direction is thus of prime scientific and technological importance: how light pumping interacts with unconventional quantum matter. 

On one hand, nontrivial solid-state matter typically entails interaction in materials, e.g., disorder or electron-phonon scattering and electron correlation, and is often theoretically challenging. 
On the other hand, non-Hermitian (NH) matter is an emerging class of effective quantum states lacking Hermiticity in their Hamiltonian\cite{Bender2007,Mostafazadeh2010,Moiseyev2009,Torres2019,Ashida2020,Bergholtz2021}. 
It has been experimentally limited to cold atomic and photonic matter and classical resonators 
typically out of equilibrium\cite{Dembowski2001,Gao2015,  Brandenbourger2019, Fruchart2021, Helbig2020, Xiao2020,  Zhang2021,XXZ:nHcircuit}. 
Combining two nonequilibrium conditions, optical pumping and non-Hermiticity, is exceptionally demanding\cite{Yuce2015,Zhou2018a,Hoeckendorf2019,Zhang2020a,Wu2020,Fedorova2020}. 
However, equilibrium solid-state systems can otherwise provide effective NH platforms: a part of the entire equilibrium system effectively acquires non-Hermiticity via interaction effects with other parts, if we look at only this part of physical degrees of freedom. 
This possibility was discussed\cite{Zyuzin2019,Papaj2019,Yoshida2018,Nagai2020} 
although experimentally hindered by unclear physical observable; recently, concrete and robust experimental observables were proposed in spin-resolved photoemission spectroscopy of a topological insulator surface state, which can reveal spin patterns induced by NH relaxation due to interaction effects\cite{Zhang2024}. 

This therefore paves the practical route towards the physics of light-NH matter hybrid, in the light of the effective NH perspective into interaction effects and establishing a unique enrichment of Floquet engineering. The surface state system has been demonstrated to host Floquet replicas via periodic THz laser pumping and angle-resolved photoemission spectroscopy (ARPES)\cite{Gedik,Gedik3,Ito2023}. Therefore, it is a suitable platform for exploring the physics of an optically pumped NH system, especially because the experimental difficulty is reduced to a normal pumping experiment of solids.
To this end, we consider shining both linear and circular polarized (LP \& CP) optical pumping fields periodically onto the topological insulator surface and measuring the ARPES signals to extract the physical information of the NH system. We reveal exotic spin soliton formation in spectroscopy to be the hallmark of this photodriven NH matter, a hybrid open system. Besides circular dichroism, intriguing topological selectiveness of different linear polarizations appears due to the intrinsically anisotropic NH relaxation. A pure pumping-induced texture shows meron strings merged to form Bloch lines (LP light) and dichroically switchable and helicity tunable skyrmions (CP light). Crucial chemical potential dependence is inspected: the $\mu_0\sim0$ case significantly distinguishes light polarizations in the pattern of the creation and annihilation of vortices in energy planes, associated with unique domain structure formation; the $\mu_0>\Omega$ case with pumping photon energy $\Omega$ bears more symmetry and a pair of intriguing 3D topological objects, a $2\pi$-twisted vortex ring related to but different from a Hopfion. We also highlight unique closed Bloch lines in the Floquet energy-momentum space.

\section*{Results}
\subsection*{Floquet NH system}\label{Sec:Keldysh}

We base on the Floquet-Keldysh formalism to study this appealing scenario. 
The two-dimensional (2D) Dirac model
\begin{equation}\label{eq:H0}
    H_0=d_\nu(\bk)\sigma^\nu= \hbar v(k_x\sigma_2 - k_y\sigma_1) + \bmm\cdot\bsigma
\end{equation}
with Pauli matrices $\sigma_\nu=(\sigma_0,\bsigma)=(I,\sigma_1,\sigma_2,\sigma_3)$ describes the surface state with exchange energy $m$ due to magnetic moment along $\hat{m}$-direction.
Since magnetic disorder typically accompanies doped magnetism along the magnetization direction $\hat{m}$\cite{lee2015,Liu2020,Tokura2019}, 
scattering anisotropy in spin is thus acquired. For example, in the first Born approximation, the matrix element of impurity scattering $V_{\sigma\sigma'}=g_0+g\sigma \delta_{\sigma\sigma'}$ for spin projection $\sigma$ along $\hat{m}$ gives rise to a disorder bath of retarded self-energy $\Sigma^\mathrm{r}_B=-\ii(\gamma_0\sigma_0+\bgamma\cdot\bsigma)$ with $\bgamma\parallel\hat{m}$ and $\gamma_0\propto g_0^2+g^2,\gamma\propto2g_0g$.
This NH relaxation $\Sigma^\mathrm{r}_B$ naturally adds to $H_0$ and leads to the NH Hamiltonian $H_0+\Sigma^\mathrm{r}_B$. 
Relevant here is in-plane $\bgamma=\gamma_1\hat{x}$ with vortex pair texture formation and broken rotation symmetry, surpassing the NH exceptional points to be the robust observables\cite{Bergholtz2021,Zhang2024}. 
Physically, one can manipulate the type and configuration of the magnetization and magnetic doping of the topological insulator surface to realize any $\bgamma$, e.g., magnetization $m\hat{x}$ entails $\gamma_1\hat{x}$.
For this in-plane magnetization, we can set $m=0$ but retain $\gamma_1$ without loss of generality since $m\hat{x}$ merely shifts the whole spin texture along $k_y$-axis and does not affect our discussion.
Pumping photon energy $\Omega$ becomes the major energy scale. This case highlights the most intriguing in-plane features and particularly enables, besides circular dichroism, a strong orthogonal LP light selection in soliton formation.

In the presence of periodic light pumping, i.e., the Floquet case, the system is stabilized in a steady state and one can rely on the photoemission formalism similar to that of an equilibrium state. The observable in spin-resolved ARPES (SARPES) measurements is the density- and spin-resolved expectation value
\begin{equation}\label{eq:rho_nu}
    \brho=\mathrm{Tr}[-\ii\bsigma G^<] 
\end{equation}
with the lesser Green's function $G^<$ defined in the Floquet-Keldysh space\cite{Rammer2011,DrivenReview} (see \textit{Methods}).
The periodic pumping vector potential 
\begin{equation}\label{eq:A(t)0}
\begin{split}
    \bA(t) = 
    \begin{cases}
        A_0\left(\delta_{\nu,1}\hat{x}\cos\frac{\Omega}{\hbar} t + \delta_{\nu,-1}\hat{y}\sin\frac{\Omega}{\hbar} t\right) & \nu=\pm1 \\
        A_0\left(\hat{x}\cos\frac{\Omega}{\hbar} t + \tau\hat{y}\sin\frac{\Omega}{\hbar} t\right) & \tau=\pm1
    \end{cases}
\end{split}
\end{equation} 
corresponds to light linearly polarized in $\hat{x}$- or $\hat{y}$-axis (XLP/YLP for $\nu=\pm1$) and right or left circularly polarized (RCP/LCP for $\tau=\pm1$) and enters Eq.~\eqref{eq:H0} via electromagnetic coupling $\hbar\bk\rightarrow\hbar\bk+e\bA(t)$.
Eq.~\eqref{eq:rho_nu} can be numerically evaluated with a cutoff in Floquet replicas.

\begin{figure*}[hbt]
\includegraphics[width=17.8cm]{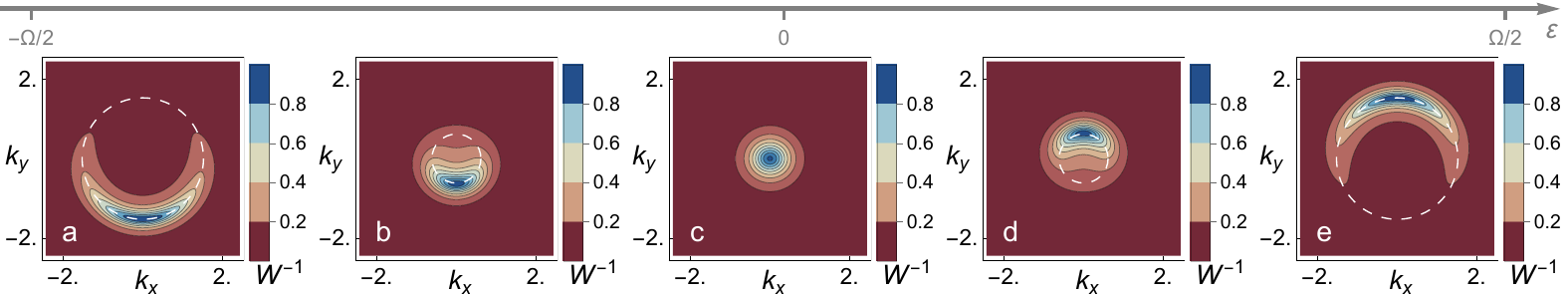}
\caption{\textbf{Energy-dependence of the overall spin magnitude profile.} Contour profile of the normalized inverse energy denominator Eq.~\eqref{eq:epsilon_denom_gamma1} in the energy-momentum space, 
which affects the spin magnitude variation the most. 
(a-e) Increasing energy planes at $\varepsilon/\Omega=-0.4,-0.2,0,0.2,0.4$ when $\gamma_0=5\mathrm{meV},\gamma_1=3\mathrm{meV},\Omega=50\mathrm{meV},v=2\times10^5\mathrm{m/s}$ and $k_{x,y}$ is measured in units of $0.01\text{\AA}^{-1}$.
The white dashed circle indicates the on-shell ring. Note that the momentum-space region with high magnitude typically has a crescent-like shape at finite energy planes. }\label{Fig:denominator_profile}
\end{figure*}

Taking the undriven case as an example and reference, Eq.~\eqref{eq:rho_nu} gives 
\begin{equation}\label{eq:rho_nu1}
    \brho=  4 f(\varepsilon)\bar \bp/W,
\end{equation}
with Fermi function $f(\varepsilon)$, energy denominator
\begin{equation}\label{eq:epsilon_denom_gamma1}
\begin{split}
    W & =   (\varepsilon^2-E^2)^2 + 4(\varepsilon\gamma_0- v\hbar k_y \gamma_1)^2
\end{split}
\end{equation}
and $E^2=d^2+(\gamma_0^2-\gamma_1^2)$.
The undriven spin texture 
\begin{equation}\label{eq:p_nu_gamma1_maintext}
\begin{split}
    \bar\bp=2(\gamma_0\varepsilon-\mycomment{\chi}v\gamma_1\hbar k_y)\bd_{12} + (\varepsilon^2-E^2) \bgamma_1
\end{split}
\end{equation}
with $\bd_{12}\equiv(d_1,d_2)$ (we henceforth use subscript `12' to denote the in-plane part of any vector). Importantly implied by the $\varepsilon$-linear term, the overall texture moves along $k_y$-axis as the energy plane varies, up to some distortion. Discussed in Supplementary Material~\ref{app:magnitude}, the spin polarization magnitude mainly controlled by Eq.~\eqref{eq:epsilon_denom_gamma1} also bears a similar $\varepsilon$-dependent movement, as shown in Fig.~\ref{Fig:denominator_profile}, where the strongest signal is roughly a \textit{crescent}-shaped blue region. 
Such $\varepsilon$-dependent movement 
will recur consequentially in driven cases.

To gain analytical insights, we develop a perturbative description in \textit{Methods} and \ref{app:lowE}. 
It provides a low-energy theory mainly for the Floquet replica $n=0$, i.e., energy $\varepsilon$ small compared to $\Omega$ and hence away from the Floquet band anticrossing points, which corroborates with full numerics. 
As the central analytical result, the effect of pumping on the photoemission spectroscopy is characterized by a low-energy Floquet texture, the driven counterpart of $\bar \bp$ in Eq.~\eqref{eq:rho_nu1},
\begin{equation}\label{eq:tildep2}
    \bp=f_0 \bar \bp+4\alpha^2 (f_0\bp' + f_1\bp^+ + f_{-1}\bp^-),
\end{equation}
where the dimensionless parameter $\alpha=e vA_0/2\Omega$
signifies the pumping strength. 
Here, Floquet-shifted Fermi occupation $f_n(\varepsilon)=f(\varepsilon-n\Omega)$ must be included to account for the contribution from different replica $n$. This consists of replica $n=0$ with pumping corrections and also the effect on replica $n=0$ from replicas $n=\pm1$. For simplicity, we henceforth focus on the case close to zero temperature only. Eq.~\eqref{eq:tildep2} is also accompanied by an energy denominator as Eq.~\eqref{eq:epsilon_denom_gamma1} and Fig.~\ref{Fig:denominator_profile} depict. Note that our formalism and discussion do not rely on or assume any smallness of the NH relaxations. Throughout the figures in the present work, we choose some typical and realistic parameters to illustrate the phenomena. However, one should keep in mind that the formulae and main physical results are not restricted by any particular parameter choice; instead, they are general for a wide range of realistic parameters as explained in the 'Discussion'. 

The first pumping correction $4\alpha^2 f_0 \bp'$ in Eq.~\eqref{eq:tildep2} incorporates the dynamically generated mass energy from CP light driving\cite{Oka2009}, i.e., the correction effectively due to $\delta m(\tau)=4\tau\mycomment{\chi}\alpha^2\Omega$
with respect to the undriven Eq.~\eqref{eq:p_nu_gamma1_maintext}. 
This is thereafter absorbed in $\bar\bp$ and discussed in \ref{app:Flo_mass_effect}. For instance, it gives rise to a \textit{dichroic} meron-antimeron pair texture induced by the CP light.
Besides, the remaining contribution satisfies $\bp^+(\nu)=\bp^-(\nu),\bp^+(\tau)=\bp^-(-\tau)$ and reads 
\begin{equation}\label{eq:p^+_gamma1_main}
\begin{split}
    \bp^+\mycomment{(\gamma_1)}=\begin{cases}
    \frac{\gamma_0\varepsilon-\nu\gamma_1d_1}{2}\bd_{12}-\frac{\nu E_+}{4}\bgamma_1+\bP_\nu & \nu=\pm1 \\
    \gamma_0( \varepsilon\bd_{12} + \gamma_0\tau\bd_{12}\times\hat{z}
    ) +\bP_\tau & \tau=\pm1
    \end{cases}
\end{split}
\end{equation} 
with $E_\pm = \varepsilon^2-d^2 \mp(\gamma_0^2-\gamma_1^2)$, $\bP_\nu=\delta_{\nu,1}\,\gamma_0\gamma_1(-\gamma_0,0,d_2)$, and $\bP_\tau=-\gamma_0\gamma_1(\gamma_0, \varepsilon\tau,\frac{\tau E_-}{2\gamma_1}- d_2)$.

We will inspect in detail a Floquet period $\varepsilon\in[-\Omega/2,\Omega/2]$ and select a few representative examples.
Our analytical 
Eqs.~\eqref{eq:tildep2}\eqref{eq:p^+_gamma1_main} play a significant role below in understanding the rich spin patterns with soliton strings, vortices, and domain structures, which are highly dependent on the energy-plane variation and the chemical potential $\mu_0$, and also distinguishes \textit{all} light polarizations in Eq.~\eqref{eq:A(t)0}.
In fact, RCP and LCP pumping textures are related by general symmetry relations given in \ref{app:symmetry}.
According to Eq.~\eqref{eq:tildep2}, when $\mu_0<0$ and comparable to $-\Omega$, mainly it is $f_{1}$ that contributes to the low-energy region not far away from $\varepsilon=0$. 
When $|\mu_0|\ll\Omega$ and thus close to zero energy, both $f_0$ and $f_1$ can contribute: for $\varepsilon>\mu_0$, $f_1$ dominates over the vanishingly small $f_0$; for $\varepsilon<\mu_0$, $\bar \bp$ from $f_0$ typically prevails as $\bp'$ and $\bp^+$ are suppressed by $\alpha^2$. 
When $\mu_0>0$ and $\mu_0\sim\Omega$ or even higher, $f_{0,\pm1}$ can all come into play.

\begin{figure}[!hbt]
\includegraphics[width=6.9cm]{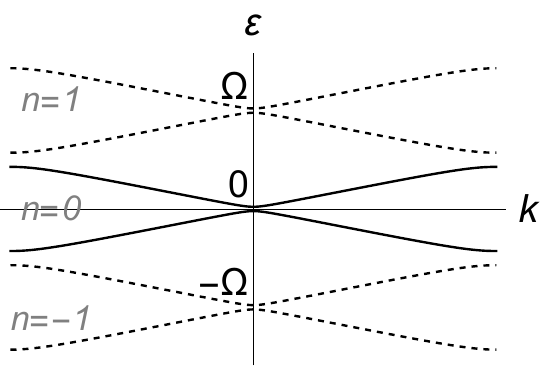}
\caption{\textbf{Schematic Floquet replica bands.} The dashed $n=\pm1$ replicas can affect the spin textures in the low-energy region of the $n=0$ replica.}\label{Fig:Floquet}
\end{figure}

The following guiding principles to understand the results will be referred to repeatedly. 
\begin{enumerate}[label=rule \roman{enumi}.,ref=rule \roman{enumi}]
    \item \label{itm:1} The perturbative low-energy theory for Floquet replica $n=0$, represented by $\bp(\varepsilon)$ from Eq.~\eqref{eq:tildep2} that partially includes effects from $n\neq0$ replicas, is most accurate for $\varepsilon,k$ small compared to $\Omega$. 
    Also, the equilibrium part $\bar\bp(\varepsilon)$ often dominates when $\varepsilon<\mu_0$ as aforementioned.
    See Fig.~\ref{Fig:Floquet} for a schematic representation of three replicas.
    \item \label{itm:2} Extra contributions are directly from $n\neq0$ replicas and thus approximately given by $\bp(\varepsilon-n\Omega)$, up to distortion due to full inter-replica hybridization. For instance, when $\mu_0\sim0$ and $\varepsilon>\mu_0$ the lower branch of the $n=1$ replica prevails over the upper branch of the $n=0$ replica, because the occupation $f(\varepsilon-\Omega)\approx1$ 
    and $f(\varepsilon)\approx0$. Finite $\mu_0$ shifts the energy range of finite contribution. 
    \item \label{itm:3} The competition between the above two determines the spin texture, where the magnitude profile of Fig.~\ref{Fig:denominator_profile} also participates. When the $n=0$ texture (\ref{itm:1}) is very different from the background texture from neighboring $n\neq0$ replicas (\ref{itm:2}), vortices and domains occur. This often well describes the behavior for $|\varepsilon|\ll\Omega/2$ and can be qualitatively extended to higher energies.
\end{enumerate}


\subsection*{Pure photodriven textures} 
\label{Sec:p^+}
Among the rich possibilities, it is noteworthy when Eq.~\eqref{eq:p^+_gamma1_main} 
is significant on its own or separable from other contributions, because it represents a \textit{pure photodriven} effect due to replica $n=1$, which is 
distinct from the undriven case. 
This is made possible when other terms 
in Eq.~\eqref{eq:tildep2} are relatively small or subtracted in the signal.
An example is the situation when $\mu_0\sim-\Omega$ and $\varepsilon<0$. Since the $n=1$ ($n=0$) replica remains filled (empty) within this region, the competition in \ref{itm:3} is mainly between $\bp^+(\varepsilon<0)$ and $\bar\bp(\varepsilon-\Omega)$ while the equilibrium part $\bar\bp(\varepsilon<0)$ is discarded. Also, $\bar\bp(\varepsilon-\Omega)$ is made as small as possible since its on-shell ring is $\Omega/v$ away from the origin $\bk=0$ of $\varepsilon=0$-plane, e.g., compared to the similar but slightly less optimal case when $\mu_0\sim0$ and $\varepsilon>0$.
Let's now inspect its spin texture. 

\subsubsection*{LP light-selective merging meron strings}\label{sec:p^+_LP}

\begin{figure}[hbt]
\includegraphics[width=8.6cm]{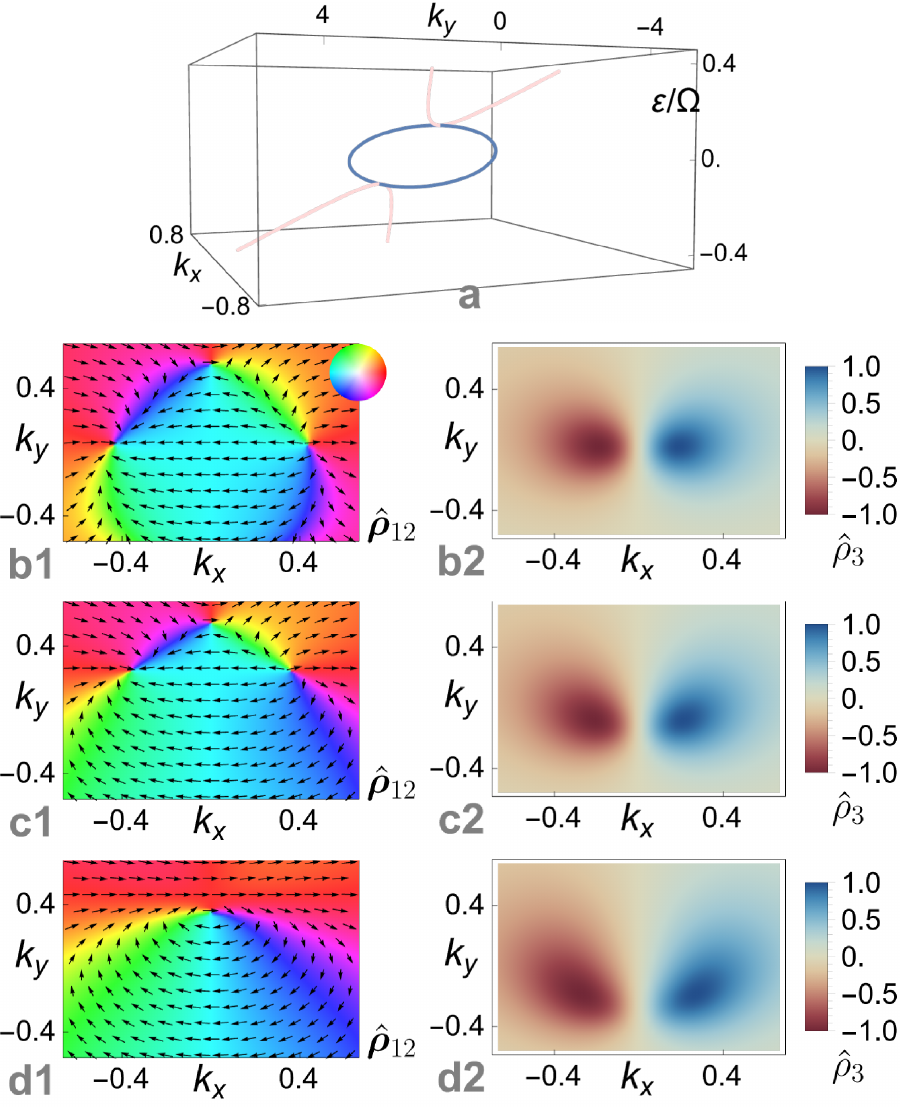}
\caption{\textbf{Pure photodriven textures under XLP light.} (a) Trajectory of the meron pair strings (dark blue) and the Bloch lines (pink) of the analytical expression of pure pumping contribution $\bp^+$. Pink Bloch lines cross dark blue meron core lines at two points. 
(b-d) Corresponding numerical data: (b1,c1,d1) in-plane normalized $\hat\brho_{12}$ and (b2,c2,d2) out-of-plane normalized $\hat\rho_3$ momentum-space spin textures under XLP light when the pure pumping effect becomes significant at decreasing energies (b,c,d) $\varepsilon/\Omega=0,-0.034,-0.07$ when $\mu_0=-50\mathrm{meV},\gamma_0=5\mathrm{meV},\gamma_1=3\mathrm{meV},A_0=7.5\times10^{-8}\mathrm{T\!\cdot\!m},\Omega=50\mathrm{meV},v=2\times10^5\mathrm{m/s}$ and $k_{x,y}$ is measured in units of $0.01\text{\AA}^{-1}$. 
In-plane spin orientation angle of $\hat\brho_{12}$, denoted by both colors and selected arrows, is plotted according to the inset of rainbow color wheel in (b). 
}\label{Fig:big_pure_tau=0}
\end{figure}

For the LP light pumping, 
note the crucial difference between $\nu=\pm1$ in Eq.~\eqref{eq:p^+_gamma1_main}. The overall spin texture profile moves \textit{oppositely}, i.e., towards the $\mp\hat{k}_y$-direction, as $\varepsilon$ increases. 
The YLP pumping has $\bp^+(\nu=-1)$ merely the same as the undriven $\bar\bp$. 
However, the XLP $\nu=1$ case is completely different as it shows pumping-induced effects (our focus below). 
Such distinction is attributed to that the NH relaxation selects $\bgamma_1$ direction in the first place.
This is our first example that the driven NH system nontrivially distinguishes two orthogonal LP lights with distinct topology; another topologically selective situation is in the $\mu_0\sim0$ case below.

For any normalized spin texture $\hat\bp(\bk)=(\cos{\Phi(\phi)}\sin{\Theta(k)},\sin{\Phi(\phi)}\sin{\Theta(k)},\cos{\Theta(k)})$ in the momentum space $\bk=k(\cos\phi,\sin\phi)$, we define the polarity, the vorticity, the skyrmion number successively as $\mathcal{P}=\frac{1}{2}\cos{\Theta(k)}\vert_{k=0}^{k=\infty}, \mathcal{V}=\frac{1}{2\pi}\Phi(\phi)\vert_{\phi=0}^{\phi=2\pi}, \mathcal{N}=\mathcal{P}\mathcal{V}$, and the helicity $\mathcal{H}=\Phi(\phi=0)$ as the in-plane winding offset.
Given these definitions, we find, in the 3D $(\bk,\varepsilon)$-space and within $-\varepsilon_0<\varepsilon<\varepsilon_0$, a pair of \textit{meron-antimeron strings} 
where 
$\varepsilon_0^2=\frac{\gamma_1^2}{\gamma_0^2-\gamma_1^2}(3\gamma_0^2+\gamma_1^2)$. 
The texture forms a ring of meron pair, shown in blue in Fig.~\ref{Fig:big_pure_tau=0}(a) and characterized by (\ref{app:p^+})
\begin{equation}\label{eq:p^+_XLP_vortexinfo1}
    \mathcal{N}=\pm1/2,\,\mathcal{P}=\pm1/2,\,\mathcal{V}=1,\,\mathcal{H}=0,\pi.
\end{equation}
The meron pair alignment herein is \textit{parallel} to $k_x$-axis, i.e., along the $\bgamma_1$-direction, in contrast to the orthogonal alignment along $k_y$-axis in the undriven 
case. 
On the other hand, outside $-\varepsilon_0<\varepsilon<\varepsilon_0$, we have a pair of \textit{Bloch lines} again merged at the $\pm\varepsilon_0$-planes, respectively in the $\varepsilon\gtrless0$ region. 
They, in each energy plane, form cores of a vortex pair characterized by vorticity and helicity
\begin{equation}\label{eq:p^+_XLP_vortexinfo2}
    \mathcal{V}=1,\,\mathcal{H}=\pm\pi/2.
\end{equation}
The above analysis and Fig.~\ref{Fig:big_pure_tau=0}(a) are for the photodriven $\bp^+$ valid mainly in the low-energy region. The $\Omega$-periodicity is restored when full Floquet bands are included as exemplified in later sections. In Fig.~\ref{Fig:big_pure_tau=0}(b-d), we indeed observe the Floquet meron-antimeron pair as we slightly lower the energy plane from $\varepsilon=0$, in which meron cores move upward and eventually merge to form a new Bloch point. The other Bloch point further away from the origin is not discernible as it is too far away from the on-shell region. 

\subsubsection*{CP light pumping: tunable skyrmion strings}

\begin{figure}[hbt]
\includegraphics[width=8.6cm]{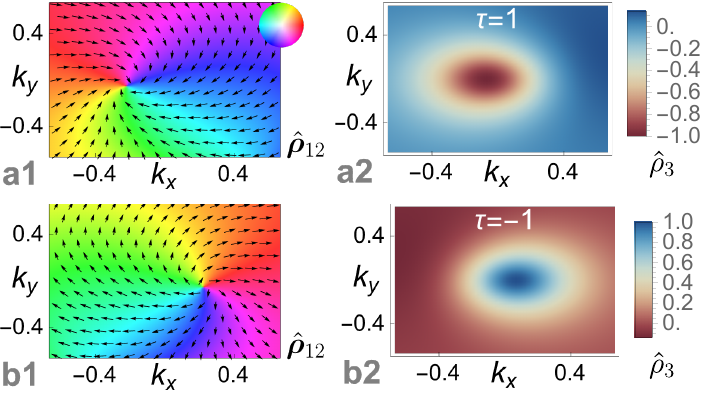}
\caption{\textbf{Pure photodriven textures under CP light.} (a1,b1) In-plane normalized $\hat\brho_{12}$ and (a2,b2) out-of-plane normalized $\hat\rho_3$ momentum-space spin textures under (a) RCP light and (b) LCP light ($\tau=\pm1$) when the pure pumping effect $\bp^+$ becomes significant. Dichroically switchable N\'eel-type skyrmion at $\varepsilon=0$-plane is exemplified. Other parameters and presentation styles follow Fig.~\ref{Fig:big_pure_tau=0}.}\label{Fig:pure_tau=pm1}
\end{figure}

\begin{figure*}[hbt]
\includegraphics[width=17.8cm]{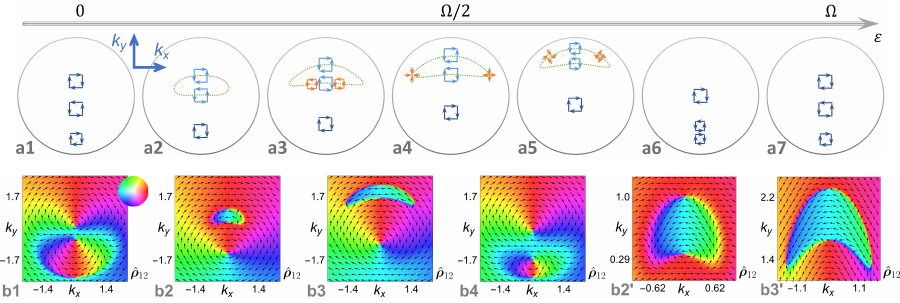}
\caption{\textbf{Energy-evolution of XLP light-driven textures for low-energy chemical potential $\mu_0$.} (a) In-plane momentum-space spin texture schematics highlight the energy-plane evolution of vortices and domains for XLP light when $\mu_0\sim0$. 
Four spins surrounding a core denote a vortex. Smaller size indicates closer to the creation or annihilation. 
Colors identify the same set of vortices. 
Green dotted lines mark the domain wall structure, where the inside and outside spins undergo a rapid change and become singular when a vortex is present.
(b) Corresponding numerical data of in-plane 
spin textures.
Selected energy planes (b1-4) $\varepsilon/\Omega=-0.03,0.18,0.5,0.87$ roughly correspond to (a1,a2,a4,a6), respectively. 
Panels (b2$^\prime$,b3$^\prime$) zoom in the domain region in (b2,b3). 
Parameters are $\gamma_0=5\mathrm{meV},\gamma_1=3\mathrm{meV},\mu_0=0,A_0=5.0\times10^{-8}\mathrm{T\!\cdot\!m},\Omega=50\mathrm{meV},v=2\times10^5\mathrm{m/s}$ and $k_{x,y}$ measured in units of $0.01\text{\AA}^{-1}$. Other presentation styles follow Fig.~\ref{Fig:big_pure_tau=0}.
}\label{Fig:big_mu0_tau=0}
\end{figure*}

For the CP light pumping ($\tau=\pm1$), Eq.~\eqref{eq:p^+_gamma1_main} gives a \textit{skyrmion string} along the $\varepsilon$-direction and through $\bK=(-\tau\gamma_1,0)$ in the $(\bk,\varepsilon)$-space, which is characterized by
\begin{equation}\label{eq:p^+_CP_vortexinfo}
    \mathcal{N}=-\tau,\,\mathcal{P}=-\tau,\,\mathcal{V}=1,\,\mathcal{H}=\phi_0(\tau,\varepsilon)
\end{equation}
with $\phi_0=\arctan{(-\gamma_0\tau,\varepsilon)}$.
Finite $\gamma_0$ assists the skyrmion string formation while 
$\gamma_1$ displaces the skyrmion core in $\bk$-space. 
Eq.~\eqref{eq:p^+_CP_vortexinfo} realizes \textit{dichroic} switching of the skyrmion number $\mathcal{N}$. 
Also, reversing energy $\varepsilon$, one only reverses the helicity $\mathcal{H}(-\varepsilon)= -\mathcal{H}(\varepsilon)$. At the special plane of $\varepsilon=0$, we have $\mathcal{H}(\tau=\pm1)=\pi,0$, which is a N\'eel-type skyrmion. Along the entire $\varepsilon$-axis from negative to positive, we have $\mathcal{H}(\tau=1)$ decreasing from $\frac{3\pi}{2}$ to $\frac{\pi}{2}$ or $\mathcal{H}(\tau=-1)$ increasing from $-\frac{\pi}{2}$ to $\frac{\pi}{2}$, i.e., intermediate skyrmions between the N\'eel-type and Bloch-type in most $\varepsilon$-planes. 
In Fig.~\ref{Fig:pure_tau=pm1}, we exemplify the dichroically switchable N\'eel-type skyrmion at $\varepsilon=0$-plane, in which the location of skyrmion core, the polarity and skyrmion number are indeed controlled by $\tau$.

\subsection*{\texorpdfstring{$\mu_0\sim0$}{Lmu} case: domain accompanying vortices and LP light-selectiveness}\label{Sec:mu=0}

Now we turn to $\mu_0\sim0$, which becomes more complicated due to multiple significant contributing factors. The main feature is vortice and domain structures formed by LP and CP light pumping. Importantly, besides the foregoing pure pumping case, 
this is the second example that our driven NH system drastically distinguishes two orthogonal LP lights and exhibits topological selectiveness: only the XLP case can induce the nontrivial textures. We thus focus here on it as a representative. The CP light cases exhibit dichroically tilted 'open domain' formation as discussed in \ref{app:mu=0}.
In Fig.~\ref{Fig:big_mu0_tau=0}(a), we summarize the main features by merely representing the vortices and possible domain structures, which serve as guidance. The numerical data are shown in Fig.~\ref{Fig:big_mu0_tau=0}(b). 
We draw Fig.~\ref{Fig:big_mu0_tau=0}(a) along the energy range from $0$ to $\Omega$. Floquet periodicity $\varepsilon\equiv \varepsilon+\Omega\pmod \Omega$ 
readily connects the figures presented to low-energy regions, e.g., $|\varepsilon|<\Omega/2$.
    
We exemplify 
$0<\varepsilon<\Omega/2$ where the equilibrium $\bar\bp(\varepsilon)$ is suppressed due to the occupation effect.
Firstly for \ref{itm:1}, at large-$(\bk,\varepsilon)$ region close to the on-shell band dispersion, Eq.~\eqref{eq:p^+_gamma1_main} 
becomes 
\begin{equation}\label{eq:p^+_gamma1_on_main}
\begin{split}
    \bp^+_\mathrm{on}(\nu)=
    \frac{1}{2}(\gamma_0\varepsilon-\nu\gamma_1d_1)\bd_{12}.
\end{split}
\end{equation}
Secondly, \ref{itm:2} can provide another background vortex winding texture roughly given by $\bar\bp(\varepsilon-\Omega)$. 
Its $\varepsilon$-dependent movement leads to 
background spins in gross pointing around $+\hat{x}$-direction. 
This exactly corresponds to the reddish region in Fig.~\ref{Fig:big_mu0_tau=0}(b2-4). 


Similar to Eq.~\eqref{eq:p^+_gamma1_main}, Eq.~\eqref{eq:p^+_gamma1_on_main} has the $\varepsilon$-dependent movement direction controlled by $\nu$. This crucially leads to $|\bp^+_\mathrm{on}(\nu=1)|\gg|\bp^+_\mathrm{on}(\nu=-1)|$ for the on-shell crescent region (see Fig.~\ref{Fig:denominator_profile}(d,e) and \ref{app:mu=0}).
Given the slowly varying background texture, 
the $\nu=1$ crescent region is much more capable of inducing interesting structures according to \ref{itm:3}. Indeed, this causes that \textit{only} the XLP ($\nu=1$) case shown in Fig.~\ref{Fig:big_mu0_tau=0} sees the persistent appearance of an extra domain with boundary vortices, which extends from low energies $\varepsilon\sim0$ even up to the highest energies $\varepsilon\sim\Omega/2$. 

\begin{figure*}[hbt]
\includegraphics[width=17.8cm]{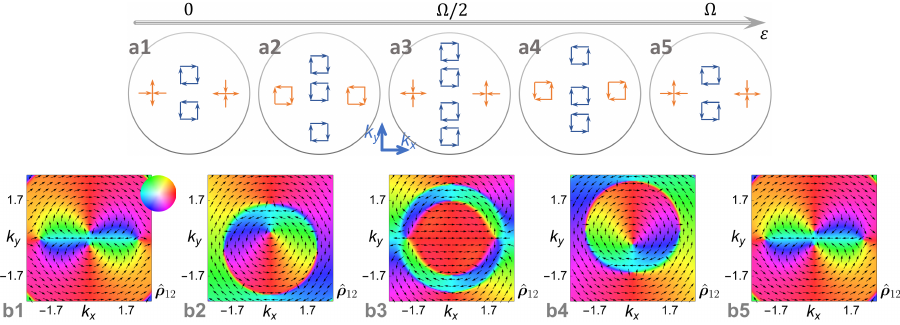}
\caption{\textbf{Energy-evolution of LP/CP light-driven textures for high-energy chemical potential $\mu_0$.} (a) In-plane momentum-space spin texture schematics highlighting the energy-dependent evolution of vortices when $\mu_0>\Omega$. Various light polarizations do not change the features qualitatively. (b) Corresponding numerical data: in-plane momentum-space spin textures under XLP light; 
CP light cases only modify the in-plane texture in a minor way and are omitted. Selected energy planes (b1-5) $\varepsilon/\Omega=0,0.2,0.5,0.8,1.0$ roughly correspond to (a1-5). Parameters are $\gamma_0=5\mathrm{meV},\gamma_1=3\mathrm{meV},\mu_0=100\mathrm{meV},A_0=2.5\times10^{-8}\mathrm{T\!\cdot\!m},\Omega=50\mathrm{meV},v=2\times10^5\mathrm{m/s},\nu=1$ and $k_{x,y}$ measured in units of $0.01\text{\AA}^{-1}$. Other presentation styles follow Fig.~\ref{Fig:big_mu0_tau=0}.
}\label{Fig:big_muH_tau=0}
\end{figure*}

Despite distortion and movement, the overall texture of such a domain follows Eq.~\eqref{eq:p^+_gamma1_on_main}, i.e., a simple vortex winding when $\varepsilon>0$. At the boundary of the crescent domain, a domain wall naturally forms to connect smoothly to the background texture. However, such a smooth connection is obstructed at two points where the $k_y$-axis crosses the domain, because the background $+\hat{x}$-spin has to directly turn to $-\hat{x}$ without pointing transversely; hence, there forms the light blue pair of in-plane vortices in Fig.~\ref{Fig:big_mu0_tau=0}(a2-4) and the corresponding Fig.~\ref{Fig:big_mu0_tau=0}(b2,b3). 
This vortex pair is directly captured by combining 
$\bp^+$ and the background $\bar\bp(\varepsilon-\Omega)$ 
\begin{equation}\label{eq:bp_mu0_main}
\begin{split}
    \bp(\bk,\varepsilon) = \frac{4\alpha^2\bp^+(\varepsilon,\nu=1)}{W(\varepsilon)} + \frac{\bar\bp(\varepsilon-\Omega)}{W(\varepsilon-\Omega)} ,
\end{split}
\end{equation}
where we include the energy denominator Eq.~\eqref{eq:epsilon_denom_gamma1}. Besides, the vortex cores are also Bloch points as discussed later for the appearance of Bloch lines.

Furthermore, 
spin textures at larger $|\varepsilon|$ less follow the low-energy analytical formulae; the region $|\varepsilon|\sim\Omega/2$ bears intrinsically high-energy and nonperturbative properties, where vortices or domain walls can persist and have extra transformation.
The other left/right orange pair of in-plane vortices in Fig.~\ref{Fig:big_mu0_tau=0}(a3-5) when $\varepsilon\sim\Omega/2$ is beyond the reach of low-energy theory, which is exemplified in Fig.~\ref{Fig:big_mu0_tau=0}(b3'). Interestingly, its trajectory, from creation to annihilation, follows the crescent domain boundary, which also abides by the general symmetry relation. 

\subsection*{\texorpdfstring{$\mu_0>\Omega$}{Hmu} case: topologically twisted vortex rings}\label{Sec:muH}

We turn to yet another set of phenomena when $\mu_0>\Omega$, which becomes simpler than the previous $\mu_0\sim0$ case, 
thanks to the high occupation that mainly singles out undriven contributions from different replicas, which thus naturally suppresses $\tau$-dependence. 
It accordingly bears the most symmetry. 
In Fig.~\ref{Fig:big_muH_tau=0}(a), we summarize the main features by representing the vortices in the absence of domain structures. The numerical data are shown in Fig.~\ref{Fig:big_muH_tau=0}(b) for the LP light only without loss of generality. 

This case has two branches of each replica in Fig.~\ref{Fig:Floquet} equally contributing, 
i.e., Eq.~\eqref{eq:tildep2} becomes $\bp= \bar\bp+4\alpha^2(\bp'+\bp^++\bp^-)$ and hence $\bar\bp$ dominates. Following \ref{itm:3}, for $|\varepsilon|\leq\Omega/2$, the texture is mainly contributed by replicas $n=0,\pm1$; indeed, 
\begin{equation}\label{eq:bp_Hmu}
    \bp(\varepsilon)=\bar\bp(\varepsilon)+\bar\bp(\varepsilon-\Omega)+\bar\bp(\varepsilon+\Omega)
\end{equation}
with energy denominator understood can well reproduce the numerics at \textit{every} $\varepsilon$-plane, not limited to low energies. 
For instance, in the $\varepsilon=0$-plane two in-plane vortices centered at $\bK^\pm=(\pm[\Omega^2/3-(\gamma_0^2-\gamma_1^2)]^\frac{1}{2},0)$
are the orange vortex pair in Fig.~\ref{Fig:big_muH_tau=0}(a1) and the corresponding Fig.~\ref{Fig:big_muH_tau=0}(b1). They are respectively characterized by 
\begin{equation}\label{eq:muH_vortexinfo1}
    \mathcal{V}=-1,\,\mathcal{H}^\pm=0,\pi.
\end{equation}
Similarly, the $\varepsilon=\Omega/2$-plane bears two orange in-plane vortices in Fig.~\ref{Fig:big_muH_tau=0}(a3), characterized by 
\begin{equation}\label{eq:muH_vortexinfo2}
    \mathcal{V}=-1,\,\mathcal{H}^\pm=\pi,0.
\end{equation}

Combining Eqs.~\eqref{eq:muH_vortexinfo1}\eqref{eq:muH_vortexinfo2}, we realize a remarkable property of the orange vortex pair in Fig.~\ref{Fig:big_muH_tau=0}(a). Within a Floquet period $\varepsilon\in[0,\Omega]$, the helicity continuously winds $\mp2\pi$ for the $\bK^\pm$ vortex. Each $2\pi$-twisted vortex string, compactified along $\varepsilon$-axis by naturally identifying $\varepsilon=0$ and $\varepsilon=\Omega$ as per the Floquet period, becomes an intriguing 3D topological twisted ring with a topological number $\mathcal{T}$, the rounds of such twist, in addition to the vortex winding number as vorticity
\begin{equation}\label{eq:muH_vortexinfotwist}
    \mathcal{V}=-1,\,\mathcal{T}^\pm=\mp1.
\end{equation}
Its difference from a Hopfion of homotopy group $\pi_3(S^2)=\mathbb{Z}$, e.g., a compactified $\pm2\pi$-twisted skyrmion string, lies in that $p_3$ herein does not possess the skyrmion-type structure\cite{Sk:review4}. Hence, our twisting topological number $\mathcal{T}=\mp1$ can also be interpreted as a linking number between any isospin trajectories in the $(\bk,\varepsilon)$-space\cite{Wilczek1983}. Note that those trajectories are closed loops due to Floquet periodicity. Therefore, their linking can be given by the well-known Gauss linking number.
Indeed, Fig.~\ref{Fig:big_muH_tau=0}(b) numerically confirms this picture. 

\subsection*{Appearance of Bloch lines}\label{Sec:BLs}

\begin{figure}[hbt]
\includegraphics[width=8.6cm]{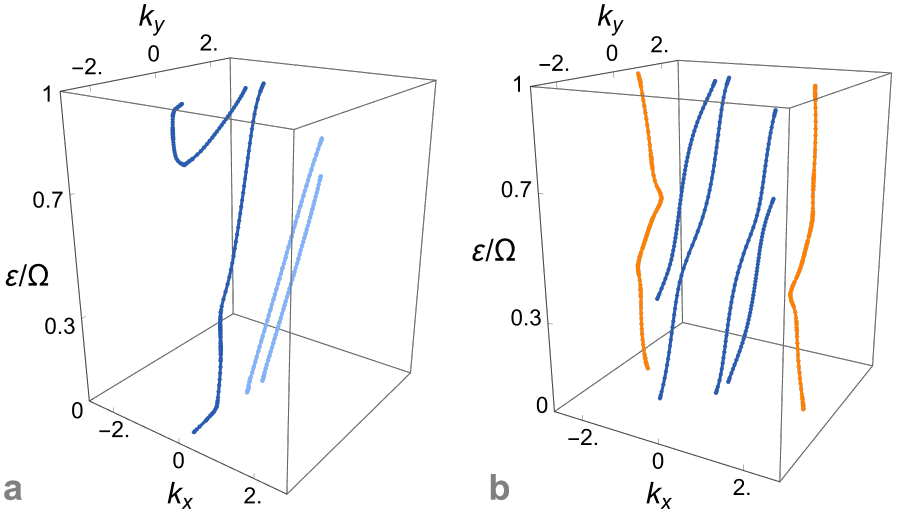}
\caption{\textbf{Bloch lines in the 3D $(\bk,\varepsilon)$-space.} $\varepsilon\in[0,\Omega]$ ranges through one Floquet period. Parameters and presentation styles follow (a) XLP light pumping for the $\mu_0\sim0$ case of Fig.~\ref{Fig:big_mu0_tau=0} and (b) XLP light pumping for the $\mu_0>\Omega$ case of Fig.~\ref{Fig:big_muH_tau=0}. Colors correspond to the schematics Fig.~\ref{Fig:big_mu0_tau=0}(a) and Fig.~\ref{Fig:big_muH_tau=0}(a), respectively.
}\label{Fig:Blochlines}
\end{figure}

Here, we inspect the appearance of Bloch lines in the 3D $(\bk,\varepsilon)$-space. They are trajectories of Bloch points where the spin magnitude vanishes, which is \textit{not} to be confused with the line core of in-plane vortices where only $\bp_{12}=\bzero$.  
We have already seen the pink Bloch lines in Fig.~\ref{Fig:big_pure_tau=0}(a) for the pure photodriven texture. 
Below, as shown in Fig.~\ref{Fig:Blochlines}, we focus on closed Bloch lines respectively corresponding to (a) XLP light pumping when $\mu_0\sim0$ of Fig.~\ref{Fig:big_mu0_tau=0} and (b) XLP light pumping when $\mu_0>\Omega$ of Fig.~\ref{Fig:big_muH_tau=0}. 
The reason why only LP light leads to Bloch lines can be intuitively understood: CP light is in general possible to transfer the photon angular momentum to the out-of-plane spin $S_z$, which can be also seen from the foregoing dynamical mass generation of $\delta m(\tau)$. 
Bloch lines are yet another interesting feature brought about by the light pumping and NH effect, and they have rarely been addressed in the energy-momentum space. For instance, an intriguing situation is the movement and crossing of closed Bloch lines in the Floquet energy-momentum space, which is the counterpart of braiding open flux lines and may generate linking structures and topologically nontrivial knots\cite{Adams2010}. We leave this possibility for future studies.

From the color correspondence between Fig.~\ref{Fig:Blochlines}(a) and Fig.~\ref{Fig:big_mu0_tau=0}(a), we see that the pair of light blue vortices, the cores of which are also Bloch points, is created not far away from $\bk=0$, moves in $\hat{k}_y$-direction, and eventually exceeds the momentum observation window. Note that the cores of high-energy orange in-plane vortices in Fig.~\ref{Fig:big_mu0_tau=0}(a3-5) are not Bloch points. 
On the other hand, the other dark blue vortices and associated Bloch lines bear more complex trajectories. One vortex comes into play from the negative $k_y$-axis and keeps moving to the positive side as $\varepsilon$ increases; another vortex-antivortex pair is, however, created at a certain energy plane $\varepsilon>\Omega/2$ (see Fig.~\ref{Fig:big_mu0_tau=0}(a6)), leading to the slanted U-shaped Bloch line in Fig.~\ref{Fig:Blochlines}(a). Due to Floquet periodicity, the much longer dark blue Bloch line is actually connected near $\varepsilon\equiv 0\pmod \Omega$ with one thread of the U-shaped Bloch line, i.e., the one near $k_y=-0.5$. Around this $\varepsilon\approx0$ region, the nearly flat part roughly along $k_y$-axis of the dark blue long Bloch line manifests the steep variation in occupation with chemical potential $\mu_0=0$. 

Similarly, Fig.~\ref{Fig:Blochlines}(b), in comparison to Fig.~\ref{Fig:big_muH_tau=0}(a), exhibits two sets of vortices with Bloch lines. Firstly, the dark blue ones correspond to the middle vortices that move in the $\hat{k}_y$-direction as $\varepsilon$ increases, during which the number of vortices varies in the manner 2-3-4-3-2. Secondly, the left and right orange ones directly correspond to the $2\pi$-twisted vortex rings in Eq.~\eqref{eq:muH_vortexinfotwist}.

\section*{Discussion}\label{Sec:discussion}

As aforementioned, the experimental detection in mind is SARPES measurement performed on the periodically pumped surface state. Both the pump and probe parts are within the reach of contemporary spectroscopy techniques. For instance, ARPES measurement showing Floquet bands 
has succeeded since a decade ago and the pump pulse in general tunes widely from THz to mid-infrared frequencies\cite{Gedik,Gedik3,Zhou2023,Ito2023}; probe pulse from synchrotron and laser source with high photon flux and long duration provides energy and momentum resolution  
down to $\lesssim1\mathrm{meV}$ and $0.005\textrm{-}0.01\mathrm{\text{\AA}^{-1}}$\cite{Jozwiak2013,Reimann2018,Lv2019,Sobota2021,Bao2021}; even time-resolved SARPES measurement of surface state continues revealing spin dynamics and resonance phenomena\cite{Cacho2015,Jozwiak2016,Iyer2018}. Note that time resolution is \textit{not} required in the present steady-state proposal, largely reducing the technical complexity including 
the compromise between time and energy resolution.

Our study mainly focuses on the regime when the dimensionless pumping strength $\alpha<0.2$ is not too large to invalidate the low-energy predictions. Indeed, the numerics can go beyond and eventually induce different excitation patterns. It is, however, not of prime importance for two reasons. i) Higher pumping fields induce quadratically more heating effects and can be destructive to the sample. ii) As detailed in \ref{app:lowE}, existing pumping and/or Floquet experiments are often estimated to be well within the leading low-energy response with $\alpha<0.1$, justifying the experimental relevance of the regime we cover in detail\cite{Cacho2015,Jozwiak2016,Zhou2023,Zhou2023a}.
To connect with the present spin-channel phenomena, we introduce a few characteristic scales. The typical energy $\varepsilon'$ is related to the finite energy plane of observation away from the band midpoint and is fundamentally related to the driving photon energy $\Omega$ as per the Floquet periodicity. From the foregoing experiments, the typical $\Omega$ can be taken as $40\textrm{-}500\mathrm{meV}$. 
From the spin dynamics in surface state pumping measurements, the spin relaxation time, the time scale of spin-resolved excitations decaying out, is at the order $0.5\textrm{-}10\mathrm{ps}$ and hence the characteristic relaxation strength $\gamma'\sim 0.4\textrm{-}8.3\mathrm{meV}$\cite{Cacho2015,Iyer2018}.
The typical momentum scale $k'$ 
can be estimated as the vortex pair separation measured between two centers $k'=\varepsilon'/\hbar v\sim 0.032\textrm{-}0.4\mathrm{\text{\AA}^{-1}}$ with Fermi velocity $v\sim2\times10^5\mathrm{m/s}$. Only for special energy planes close to vortex creation or annihilation, the separation scale might be challenging to resolve. Overall, these estimations indicate that the proposed phenomena fall well within the feasible range of detection.

The integration of NH effects and optical engineering also offers opportunities for applications that leverage their spin textures and light-matter coupling\cite{Oka2019,Harper2020,Rudner2020}. 
The polarization-dependent spin textures, such as meron pairs under LP light and helicity-switchable skyrmions under CP light, could enable optoelectronic components; further, the design of new types of detection means could utilize the LP light-selectiveness where only one of two orthogonal LP lights induces topological textures. The surface states thus may serve as active layers in polarization-sensitive photodetectors or optical modulators, where the light-NH matter coupling helps enhancing contrast in the design. Additionally, the chemical potential dependence of vortex/domain formation suggests a route to electrically tunable spin textures. Gating the surface could switch between distinct topological configurations and hence encode data in their spin structure, implying design of transient memory elements operated under optical pumping. 
The sensitivity of NH spin textures to magnetic disorder also suggests a possibility for surface defect detection, relevant for quality control in heterostructures based on topological insulators. Shifts and distortion in vortex rings/Bloch lines could map impurity distributions with nanoscale resolution, which may also depend on the advancement of techniques like nano-ARPES\cite{Lv2019,Sobota2021}. 
Despite these examples grounded in the physical properties of the present system and existing experimental capabilities, we emphasize that the present study focuses on the side of fundamental physics and describing this light-NH matter hybrid system.

\section*{Conclusion}
Light-NH matter coupled physics has remained largely inaccessible due to complex nonequilibrium. We propose that it be studied on a solid-state NH platform of topological insulator surface state with disorder/interaction induced relaxation. From the angle of NH systems, it extends the Floquet optical engineering towards the direction with interactions of electrons. Targeted at observable effects in spin-resolved photoemission spectroscopy, we find plenty of tunable topological spin textures and their delicate dependence on energy, chemical potential, and light polarization, which are only made possible by a combination of light pumping and NH relaxation. The results reveal the vast potential of light-matter hybridization in conjunction with open-system-like conditions, which comes from either interaction effects of interest to solid-state physics or even nonequilibrium non-Hermiticity in wider metamaterial contexts.

\section*{Methods}
\subsection*{Floquet-Keldysh formalism for driven NH system}
For generic nonequilibrium systems, one can define a set of double-time Green's functions\cite{Rammer2011}. They are compactly given as $\hat{G}=\begin{bmatrix}
G^t & G^< \\
G^> & G^{\bar{t}}
\end{bmatrix}$, where, for electronic annihilation/creation operator $c_a^{(\dag)}$ of an arbitrary quantum number labelled by $a$, the time-ordered $G^t(a,t_1;b,t_2)=-\ii/\hbar\braket{\cT c_a(t_1)c_b^\dag(t_2)}$, the similar but anti-time-ordered $G^{\bar{t}}(a,t_1;b,t_2)=-\ii/\hbar\braket{\bar\cT c_a(t_1)c_b^\dag(t_2)}$, the lesser $G^<(a,t_1;b,t_2)=\ii/\hbar\braket{c_b^\dag(t_2)c_a(t_1)}$ and the greater $G^>(a,t_1;b,t_2)=-\ii/\hbar\braket{c_a(t_1)c_b^\dag(t_2)}$. One can perform a transformation $\check{G}=L\tau_3\hat{G}L^\dag$ with $L=(1-\ii\tau_2)/\sqrt{2}$ and the Pauli matrices $\tau_\mu,\mu=0,\cdots,3$ in the $2\times2$ Keldysh space of $\hat{G}$ and $\check{G}$.
Then the transformed Green's function and self-energy share the same structure
$\check{G}=\begin{bmatrix}
G^\mathrm{r} & G^\mathrm{k} \\
0 & G^\mathrm{a}
\end{bmatrix}$
and 
$\Sigma_B=\begin{bmatrix}
\Sigma^\mathrm{r} & \Sigma^\mathrm{k} \\
0 & \Sigma^\mathrm{a}
\end{bmatrix}$ where $B$ denotes the NH relaxation bath;
they also satisfy the same relation to the lesser one $G^<=\frac{1}{2}(G^\mathrm{a} - G^\mathrm{r}  +G^\mathrm{k}),\Sigma^<=\frac{1}{2}(\Sigma^\mathrm{a} - \Sigma^\mathrm{r}  +\Sigma^\mathrm{k})$. 
For the steady state, the lesser Green's function satisfies $G^< = G^\mathrm{r} \Sigma_B^< G^\mathrm{a}$. In numerical calculations, we can thus mainly rely on the transformed $\check{G}$ and turn back when necessary.

From Eq.~\eqref{eq:H0}, the time-dependent Hamiltonian due to the electromagnetic coupling reads
\begin{equation}
\begin{split}
H_0(t)=H_0(\hbar\bk+e\bA(t))=H_0(\bk)+e\bA(t)\cdot\bv
\end{split}
\end{equation}
where $\bv=\partial H_0/\hbar\partial\bk$ is the velocity operator. To deal with this periodic driving, we need to introduce the Floquet structure to Green's functions and self-energies\cite{DrivenReview}. For instance, we can transform $\check{G}$ as a function of energy $\omega$ to the Floquet space 
\begin{equation}
\begin{split}
    \check{G}_{mn}(\omega)=\frac{1}{T_0}\int_0^{T_0} \dd & t_a \int_{-\infty}^{\infty} \dd t_r \, \ee^{-\ii(m-n)\Omega t_a/\hbar} \\
    &\times\ee^{\ii(\omega -\frac{m+n}{2}\Omega)t_r/\hbar} \check{G}(t_a,t_r)   
\end{split}
\end{equation}
where $m,n$ are the Floquet indices, the driving period $T_0=\frac{2\hbar\pi}{\Omega}$ and $t_a=(t_1+t_2)/2,t_r=t_1-t_2$.
In this Floquet-Keldysh space, we have the concise Dyson equation
\begin{equation}\label{eq:Dyson_full}
    (\check{G}^{-1})_{mn} = 
    [\delta_{mn}(\omega-n\Omega) - H^{mn}]\tau_0 + (\Sigma_B)_{mn}.
\end{equation}
The Floquet Hamiltonian in Eq.~\eqref{eq:Dyson_full} is defined as 
\begin{equation}
\begin{split}
    H^{mn}=\frac{1}{T_0}\int_0^{T_0} H_0(t) \ee^{\ii(m-n)\Omega t}\dd t,
\end{split}
\end{equation}
which is nonzero only for the Floquet-diagonal and
the inter-replica one-photon processes $H^\pm=H^{n\pm1,n}=\frac{eA_0}{2} v^\pm$ with 
$v^\pm=\begin{cases}
    \delta_{\nu,1}v_1\pm\ii\delta_{\nu,-1} v_2 & \textrm{LP light} \\
    v_1\pm\ii\tau v_2 & \textrm{CP light}
\end{cases}$.
Considering the anisotropic impurity scattering due to magnetic doping or the wide-band coupling to a general fermionic bath\cite{Zhang2024}, one can find the relevant three self-energies specified as $(\Sigma^\mathrm{r(a)}_B)_{mn}=\mp\ii\gamma_\nu\sigma^\nu\delta_{mn}$ and $
    (\Sigma_B^<)_{mn} =  2\ii\gamma_\nu f_n\sigma^\nu\delta_{mn}$.
Here, $f_n(\omega)=f(\omega-n\Omega)$ is the Floquet-shifted Fermi distribution and $\gamma_0$ and $\bgamma$ are respectively the relaxation rate in the charge and spin channels.

Eq.~\eqref{eq:Dyson_full} forms the basis of our numerical and analytical calculations. In numerics, one can solve it up to a reasonable cutoff of high Floquet replicas.
In this exact formulation, periodic driving is fully accounted for via the infinite replicas. In contrast, in the low-energy theory one should instead include the replicas approximately through the self-energy corrections, order by order.
The physical process and approximation are diagrammatically represented below in the Keldysh space. The thin line is the bare surface state electron propagator; the thick line is the electron dressed by NH relaxation from bath $\Sigma_B$ due to interaction effects such as disorder scattering; the double line is the electron further influenced by two neighboring Floquet replicas via the wavy photon line:
\begin{equation}\label{eq:diagram}
\begin{split}
\vcenter{\hbox{\feynmandiagram [inline=(a.base), horizontal=a to b, layered layout] {
  a -- [fermion, thick] b 
};}}
&\,=\,
\vcenter{\hbox{\feynmandiagram [inline=(a.base), horizontal=a to b, layered layout] {
  a -- [fermion] b 
};}}
\,+\,
\vcenter{\hbox{\begin{tikzpicture}[baseline=(a.base)]
\begin{feynman}[inline=(a.base)]
\vertex (a);
\node[right=0.6cm of a,draw,fill=white,rectangle] (b){$\Sigma_B$}; 
\vertex[right=1cm of b] (c);
\diagram*[horizontal=(a) to (c)] 
{(a)  -- [fermion] (b), (b) -- [fermion, thick] (c) 
};
\end{feynman}
\end{tikzpicture}}}\\
\vcenter{\hbox{\feynmandiagram [inline=(a.base), horizontal=a to b, layered layout] {
  a -- [double, with arrow=0.5, edge label=\(n\)] b 
};}}
\,\approx\,
\vcenter{\hbox{\feynmandiagram [inline=(a.base), horizontal=a to b, layered layout] {
  a -- [fermion, thick, edge label=\(n\)] b 
};}}
&\,+\,
\vcenter{\hbox{\begin{tikzpicture}[baseline=(a1.base)]
\begin{feynman}[inline=a1.base)]
\vertex (a1);
\vertex[right=0.723cm of a1] (a2);
\vertex[right=0.723cm of a2] (a3);
\vertex[right=0.723cm of a3] (a4);
\diagram* {
{
(a1) -- [fermion, thick, edge label=\(n\)] (a2) -- [fermion, thick, edge label=\(n\!+\!1\)] (a3) -- [double, with arrow=0.5, edge label=\(n\)] (a4),
},
(a2) -- [photon, out=-90, in=-90, looseness=1.0] (a3)
};
\end{feynman}
\end{tikzpicture}}}
\,+\,
\vcenter{\hbox{\begin{tikzpicture}[baseline=(a1.base)]
\begin{feynman}[inline=a1.base)]
\vertex (a1);
\vertex[right=0.723cm of a1] (a2);
\vertex[right=0.723cm of a2] (a3);
\vertex[right=0.723cm of a3] (a4);
\diagram* {
{
(a1) -- [fermion, thick, edge label=\(n\)] (a2) -- [fermion, thick, edge label=\(n\!-\!1\)] (a3) -- [double, with arrow=0.5, edge label=\(n\)] (a4),
},
(a2) -- [photon, out=-90, in=-90, looseness=1.0] (a3)
};
\end{feynman}
\end{tikzpicture}}}
\end{split}
\end{equation}







\twocolumngrid
\bibliography{reference.bib}  
\let\addcontentsline\oldaddcontentsline

\clearpage
\onecolumngrid
\newpage
{
	\center \bf \large 
	Supplemental Information \\
	\large for ``\newtitle"\vspace*{0.1cm}\\ 
	\vspace*{0.5cm}
}
\begin{center}
	Xiao-Xiao Zhang$^{1}$ and Naoto Nagaosa$^{2,3}$\\
	\vspace*{0.15cm}
    \small{$^1$\textit{Wuhan National High Magnetic Field Center and School of Physics, Huazhong University of Science and Technology, Wuhan 430074, China}}\\
	\small{$^2$\textit{RIKEN Center for Emergent Matter Science (CEMS), Wako, Saitama 351-0198, Japan}}\\
    \small{$^3$\textit{Fundamental Quantum Science Program, TRIP Headquarters, RIKEN, Wako, Saitama 351-0198, Japan}}
	\vspace*{0.25cm}	
\end{center}

\twocolumngrid	

\tableofcontents


\setcounter{section}{0}
\setcounter{equation}{0}
\setcounter{figure}{0}
\setcounter{table}{0}
\setcounter{page}{1}
\renewcommand{\theequation}{S\arabic{equation}}
\renewcommand{\thefigure}{S\arabic{figure}}
\renewcommand{\thetable}{S\arabic{table}}
\renewcommand{\theHtable}{Supplement.\thetable}
\renewcommand{\theHfigure}{Supplement.\thefigure}
\renewcommand{\thesection}{Supplementary Note~\arabic{section}}
\renewcommand{\bibnumfmt}[1]{[S#1]}
\renewcommand{\citenumfont}[1]{S#1}




\section{Driven NH system formulation and perturbation theory}
\subsection{General formula and equilibrium case}\label{app:identity}

In order to find the spin texture from Eq.~\eqref{eq:rho_nu}, it is analytically helpful to derive the following general formula after the requisite algebraic manipulation
\begin{equation}\label{eq:r<a}
\begin{split}
    & (\tilde{\varepsilon}\sigma_0-\tilde{d}\cdot\bsigma)^{-1} (\zeta_\nu\sigma^\nu ) (\tilde{\varepsilon}^*\sigma_0-\tilde{d}^*\cdot\bsigma)^{-1} = \frac{\zeta_\nu \cA^\nu(\tilde{\lambda})}{|\tilde{\varepsilon}^2-\tilde{d}^2|^2} 
\end{split}
\end{equation}
where $\tilde{\varepsilon}=\varepsilon+\ii\varepsilon'$, $\tilde{\bd}=\bd+\ii\bd'$,  $\varepsilon=\omega-d_0,\varepsilon'=\gamma_0,\bd'=-\bgamma$ apparently serves for the equilibrium case but can be readily generalized shortly.
The 4-vector $\cA^\nu(\tilde{\lambda})$ is given by 
\begin{equation}\label{eq:cA0}
\begin{split}
    \cA_0 &=a_0^+\sigma_0 + \ba^+\cdot\bsigma \\
    \bcA &= \ba^-\sigma_0 + a_0^-\bsigma +  \dvec{t}\cdot\bsigma - \bn\times\bsigma  
\end{split}
\end{equation}
with the notation
\begin{equation}\label{eq:cA0_coef}
\begin{split}
    a_0^\pm = |\tilde{\varepsilon}|^2 \pm |\tilde{d}|^2,\qquad &\ba^\pm=2(\varepsilon \bd + \varepsilon'\bd'\pm\bd\times\bd') \\
    \dvec{t} = 2(\bd\bd+\bd'\bd'),\qquad &\bn=2 (\varepsilon'\bd-\varepsilon\bd').
\end{split}
\end{equation}
Henceforth, we introduce shorthand 4-vector notation 
\begin{equation}\label{eq:tilde_lambda}
    \tilde{\lambda}_\mu=(\tilde{\varepsilon},\tilde{\bd})
\end{equation}
and the generic $\zeta_\nu$ in Eq.~\eqref{eq:r<a} can take different values to be specified below. 

As an example, for the equilibrium Green's functions denoted by $\cG$, we have the retarded and advanced ones
\begin{equation}
    \cG^\mathrm{r(a)}=\frac{1}{\omega-H_0-\Sigma_B^\mathrm{r(a)}} = \left( \frac{1}{\tilde{\varepsilon}\sigma_0-\tilde{d}\cdot\bsigma} \right)^{(\dag)} 
\end{equation}
and Eq.~\eqref{eq:r<a} helps us obtain 
\begin{equation}\label{eq:cG^<}
\begin{split}
    \cG^<(\omega) & =\cG^\mathrm{r} \Sigma_B^< \cG^\mathrm{a} =  \frac{2\ii f(\omega)}{W} \gamma_\nu \cA^\nu (\tilde{\lambda}),
\end{split}
\end{equation}
in which 
\begin{equation}\label{eq:epsilon_denom0}
\begin{split}
    W(\bk,\varepsilon)&=|\tilde{\varepsilon}^2-\tilde{d}^2|^2 
    = (\varepsilon^2-E^2)^2 + 4(\varepsilon\gamma_0+\bd\cdot\bgamma)^2
\end{split}
\end{equation}
with $E^2=d^2+(\gamma_0^2-\gamma^2)$.
This readily leads to the undriven texture
\begin{equation}\label{eq:barp_nu}
\begin{split}
    \bar p_0& = \gamma_0(\varepsilon^2+E^2) + 2 \varepsilon\bd\cdot\bgamma \\
    \bar\bp &= 2(\gamma_0\varepsilon+\bd\cdot\bgamma)\bd + (\varepsilon^2-E^2) \bgamma.
\end{split}
\end{equation}

\begin{widetext}
\subsection{Low-energy Floquet theory}\label{app:lowE}

In the following, we base on the Floquet-Keldysh formalism and derive the low-energy effective Green's function of replica $n=0$ Eq.~\eqref{eq:G^<1}. We use subscripts $B$ and $F$ respectively for the contribution from the NH relaxation bath and the Floquet pumping and subscript $n$ for the Floquet replica index. Applying Eqs.~\eqref{eq:rho_nu}\eqref{eq:Dyson_full},
we have the lesser Green's function
\begin{equation}\label{eq:G^<_Floquet}
\begin{split}
    G_n^<=G_n^\mathrm{r}\Sigma_n^<G_n^\mathrm{a}, \quad &\Sigma_n^<=\Sigma_{B,n}^<+\Sigma_{F,n}^<\\
    \Sigma_{B,n}^<=2\ii f_n\gamma_\nu\sigma^\nu,\quad &\Sigma_{F,n}^<=H^-\tilde{G}_{n+1}^<H^++H^+\tilde{G}_{n-1}^<H^-
\end{split}
\end{equation}
and also the causal Green's functions involved
\begin{equation}\label{eq:G^ra_Floquet}
\begin{split}
    G_n^{\mathrm{r(a)}}&=(\omega-n\Omega-H_0-\Sigma^{\mathrm{r(a)}})^{-1}, \quad \Sigma_n^{\mathrm{r(a)}}=\Sigma_{B,n}^{\mathrm{r(a)}}+\Sigma_{F,n}^{\mathrm{r(a)}}\\
    \Sigma_{B,n}^{\mathrm{r(a)}}&=\mp\ii \,\gamma_\nu\sigma^\nu,\quad \Sigma_{F,n}^{\mathrm{r(a)}}=H^-\tilde{G}_{n+1}^{\mathrm{r(a)}}H^++H^+\tilde{G}_{n-1}^{\mathrm{r(a)}}H^-.
\end{split}
\end{equation}
Due to the fact that any replica $n$ has two sides of neighbouring replica modes, we distinguish between $G_n$ and $\tilde{G}_{n}$, where the latter is the Floquet Green's function defined only for the respective semi-infinite systems, i.e., the $m\geq n$ block or the $m\leq n$ block.
\end{widetext}
Applying Eq.~\eqref{eq:diagram} to Eq.~\eqref{eq:G^<_Floquet}, we will focus on the $n=0$ replica. With the approximation from Eq.~\eqref{eq:G^ra_Floquet}
\begin{equation}\label{eq:G_Bn^ra_approx}
    G_{B,n}^\mathrm{r(a)}\approx-\frac{1}{n\Omega}-\frac{1}{(n\Omega)^2}(\tilde{\varepsilon}\sigma_0-\tilde{\bd}\cdot\sigma),
\end{equation}
valid when $\Omega$ is larger compared to the typical low-energy scale within the $n=0$ replica of interest, we immediately have 
\begin{equation}\label{}
\begin{split}
    G_{B,\pm1}^< & = G_{B,\pm1}^\mathrm{r}\Sigma_{B,\pm1}^<G_{B,\pm1}^\mathrm{a} \approx \frac{2\ii f_{\pm1}}{\Omega^2} \gamma_\nu\sigma^\nu.
\end{split}
\end{equation}
Note that the electrons with Floquet indices $n\pm1$ in the Floquet self-energy $\Sigma_{F,n}$ are not recursively dressed in the diagrammatic Eq.~\eqref{eq:diagram}, which helps us eventually ignore the foregoing difference between $G_n$ and $\tilde{G}_{n}$.
The above enables us to find, after the requisite algebraic manipulation,
\begin{equation}\label{eq:Sigma^<_F0}
\begin{split}
    \Sigma_{F,0}^< & = H^-G_{B,1}^<H^+ + H^+G_{B,-1}^<H^- \\
    &= 2\ii \:\alpha^2 [ f_{1}v^-\gamma_\nu \sigma^\nu v^+ + f_{-1}v^+\gamma_\nu \sigma^\nu v^-]\\
    &= 2\ii \:\alpha^2 \sum_{m=\pm1} f_m\Gamma_\nu^m \sigma^\nu
\end{split}
\end{equation}
with
\begin{equation}\label{eq:Gamma^pm}
\begin{split}
    \Gamma_\mu^\pm&=\left(\gamma_0(1+\tau^2)\pm2\gamma_3\tau\mycomment{\chi},-\nu\gamma_1(1-\tau^2),\right.\\
    &\left.\nu\gamma_2(1-\tau^2),\mp2\gamma_0\tau\mycomment{\chi}-\gamma_3(1+\tau^2)\right).
\end{split}
\end{equation}
On the other hand, we have from Eq.~\eqref{eq:G_Bn^ra_approx} and Eq.~\eqref{eq:G^ra_Floquet} that 
\begin{equation}\label{eq:Sigma^ra_F0}
\begin{split}
    \Sigma_{F,0}^\mathrm{r(a)} & = H^-G_{B,1}^\mathrm{r(a)}H^+ + H^+G_{B,-1}^\mathrm{r(a)}H^- \\
    &= \frac{1}{\Omega}[H^+,H^-] -2\alpha^2 \tilde\Upsilon_\nu(\tilde{\lambda}')\sigma^\nu \\
    &= \delta m\sigma_3 -2\alpha^2 \tilde\Upsilon_\nu(\tilde{\lambda}')\sigma^\nu
\end{split}
\end{equation}
where we denote $\tilde{\lambda}_\mu'=(\tilde{\varepsilon},-\tilde{\bd})$ and 
\begin{equation}\label{eq:delta_m0}
    \delta m(\tau)=4\tau\mycomment{\chi}\alpha^2\Omega
\end{equation}
is the mass energy dynamically generated. Also,
we use colon to denote the elementwise product
\begin{equation}
    \tilde\Upsilon_\nu(\tilde{\lambda}')=(\Upsilon\colonprod\tilde{\lambda}')_\nu=\Upsilon_\nu\tilde{\lambda}_\nu'
\end{equation}
with 
\begin{equation}\label{eq:Upsilon}
\begin{split}
    \Upsilon_\mu&=\left(1+\tau^2,-\nu(1-\tau^2),\nu(1-\tau^2),-(1+\tau^2)\right).
\end{split}
\end{equation}
Now, with Eq.~\eqref{eq:G^ra_Floquet} and Eq.~\eqref{eq:Sigma^ra_F0} we are ready to update $\tilde{\lambda},\tilde{\varepsilon},\tilde{\bd}$ in Eq.~\eqref{eq:tilde_lambda} respectively to $\tilde{\Lambda},\tilde{\cE},\tilde{\bcD}$ according to the relation
\begin{equation}\label{eq:tildeLambda}
\begin{split}
    \tilde{\Lambda}_\nu=(\tilde{\cE},\tilde{\bcD})=(1+2\alpha^2\Upsilon\colonprod)\tilde{\lambda}+\delta m\hat{z}.
\end{split}
\end{equation}

Repeatedly using Eq.~\eqref{eq:r<a} by setting $\zeta_\nu=\gamma_\nu$ or $\Gamma_\nu^\pm$, and together with Eqs.~\eqref{eq:G^<_Floquet}\eqref{eq:Sigma^<_F0}\eqref{eq:tildeLambda}, the diagrammatic Eq.~\eqref{eq:diagram} eventually takes the following form for the $n=0$ replica
\begin{equation}\label{eq:G^<0}
\begin{split}
    G_0^<(\varepsilon) & = \frac{2\ii}{|\tilde{\cE}^2-\tilde{\cD}^2|^2} \left[ f_0 \gamma_\nu + \alpha^2\left(  f_1\Gamma_\nu^++f_{-1}\Gamma_\nu^-\right) \right]\cA^\nu(\tilde{\Lambda}) \\
\end{split}
\end{equation}
One can simplify by neglecting in Eq.~\eqref{eq:G^<0} the cross effect between $\Sigma_{F,n}^{\mathrm{r(a)}}$ and $\Sigma_{F,n}^{<}$ that enters Eq.~\eqref{eq:G^<_Floquet}, i.e., the contribution even higher order than $O(\alpha^2)$, which leads to 
\begin{equation}\label{eq:G^<01}
\begin{split}
    G_0^<(\varepsilon) & = \frac{2\ii}{|\tilde{\cE}^2-\tilde{\cD}^2|^2} \left[ f_0 \gamma_\nu \cA^\nu(\tilde{\Lambda}) \right.\\
    &\left.+ \alpha^2\left(  f_1\Gamma_\nu^++f_{-1}\Gamma_\nu^-\right) \cA^\nu(\tilde{\lambda}) \right].
\end{split}
\end{equation}
We further note that the denominator can as well be approximated by the undriven equilibrium expression Eq.~\eqref{eq:epsilon_denom0}, leading immediately to 
\begin{equation}\label{eq:G^<1}
\begin{split}
    G_0^<(\varepsilon) & = \frac{2\ii}{W} \left[ f_0 \gamma_\nu \cA^\nu(\tilde{\Lambda}) + \alpha^2\left(  f_1\Gamma_\nu^++f_{-1}\Gamma_\nu^-\right) \cA^\nu(\tilde{\lambda}) \right].
\end{split}
\end{equation}
To see this, one should notice that the pumping effect of Eq.~\eqref{eq:tildeLambda} on the energy denominator $|\tilde{\cE}^2-\tilde{\cD}^2|$ is at most of order $O(\alpha^2)$, which is negligible as long as the equilibrium part does not vanish. It vanishes if and only if $\varepsilon^2-E^2=0$ and $\varepsilon\gamma_0+\bd\cdot\bgamma=0$ hold simultaneously. Combined, it leads to a contradiction $d^2+(\gamma_0^2-\gamma^2)=(\bd\cdot\bgamma)^2/\gamma_0^2<d^2$, provided that $\gamma<\gamma_0$.

Before closing this section, it is useful to estimate the realistic strength of optical pumping fields as per the dimensionless parameter
\begin{equation}
    \alpha=e vA_0/2\Omega
\end{equation}
given also in the main text together with Eq.~\eqref{eq:tildep2}. 
To compare with the literature with explicit electric field strength $E_0$, we note that the vector potential strength is computed by $A_0=\hbar E_0/\Omega$ and, when necessary, $E_0$ can be estimated from the energy flux density $I_0=\frac{c\epsilon_0}{2}E_0^2$ and fluence $F=I_0t$ with pulse duration $t$ used in the pumping processes.
$E_0$ is directly given as $E_0\sim6.8\times10^7\mathrm{V/m}$ with a typical photon energy of $400\mathrm{meV}$\cite{Zhou2023} and also $E_0\sim5\times10^7\mathrm{V/m}$ with a typical photon energy of $300\mathrm{meV}$\cite{Zhou2023a}. Besides, we have $F=0.05\mathrm{mJ/{cm}^2}$ with $t\sim200\mathrm{fs}$ and photon energy $1.5\mathrm{eV}$\cite{Jozwiak2016} and $F=0.5\mathrm{mJ/{cm}^2}$ with $t\sim150\mathrm{fs}$ and photon energy $1.55\mathrm{eV}$\cite{Cacho2015} for the Ti:Sa fundamental output. Then, assuming a typical Fermi velocity $v=0.4\times10^6\mathrm{m/s}$, we find $\alpha=0.05,0.07,0.0024,0.008$ respectively in the above four pumping and/or Floquet experiments. Clearly, they are all well within the range of validity $\alpha\ll1$ of our low-energy theory. In fact, numerical calculations can confirm that the phenomena of our focus remain valid in gross for $\alpha<0.25$, which goes beyond the usual expectation of perturbation theory owing to the topological robustness of the textures. These considerations relevant to realistic experiments justify the importance of our analytic analysis based on perturbation theory.

\subsection{Derivation of the spin textures}\label{app:texture}
Here, we give the derivation of the low-energy spin textures in the main text. 
In the first term of Eq.~\eqref{eq:G^<1}, we can retain only the contribution up to $O(\alpha^2)$ in $\cA(\tilde{\Lambda})$, which is entirely due to the fact that the argument $\tilde{\lambda}$, given in Eq.~\eqref{eq:tilde_lambda}, of Eq.~\eqref{eq:cA0} is replaced by $\tilde{\Lambda}$, given in Eq.~\eqref{eq:tildeLambda}. Accordingly, Eq.~\eqref{eq:cA0_coef} will be modified to 
\begin{equation}\label{eq:cA1_coef}
\begin{split}
    \tilde a_0^\pm =  a_0^\pm+4\alpha^2a_0^{\pm\prime},\qquad &\tilde\ba^\pm=\ba^\pm+ 4\alpha^2 \ba^{\pm\prime} \\
    \dvec{\tilde t} = \dvec{t} + 4\alpha^2 \;\dvec{t}',\qquad &\tilde\bn=\bn+4\alpha^2\bn'.
\end{split}
\end{equation}
And it naturally leads to the correction $\bar p_\nu+4\alpha^2 p_\nu'$ 
to the undriven spin texture in Eq.~\eqref{eq:barp_nu}, where 
\begin{equation}\label{eq:p_nu'}
\begin{split}
    p_0'&=d_0'a_0^{+\prime} - \bd'\cdot\ba^{-\prime} +\delta p_0\\
    \bp'&=d_0'\ba^{+\prime}-a_0^{-\prime}\bd' -\bn'\times\bd'-\bd'\cdot\dvec{t}' + \delta \bp.
\end{split}
\end{equation}
Here, we separate $\delta p_\nu$ purely due to the dynamically generated mass Eq.~\eqref{eq:delta_m0} in Eq.~\eqref{eq:tildeLambda}. 
Now we specify the following notation 
$\bD_\Upsilon=\Upsilon_0\bd+\bd_\Upsilon =2(d_1\tau^2,d_2,0) ,\,
    \bD_{\Upsilon\pm}'=\Upsilon_0\bd'\pm\bd_\Upsilon' =\begin{cases}
    2(d_1'\tau^2,d_2',0)  \\
    2\bd'\colonprod(1,\tau^2,1+\tau^2) 
    \end{cases},\,
    \bD_\Upsilon^\times=\bd\times\bd_\Upsilon'+\bd_\Upsilon\times\bd' =2(\bd\times\bgamma)\colonprod(\tau^2,1,0)$ 
with $\bd_\Upsilon=\bUpsilon\colonprod\bd$
and thus have the quantities in Eqs.~\eqref{eq:cA1_coef}\eqref{eq:p_nu'}
\begin{equation}\label{}
\begin{split}
    a_0^{\pm\prime}=\Upsilon_0|\tilde\varepsilon|^2 \pm \Upsilon^i|\tilde d_i|^2,\quad
    \ba^{\pm\prime}=\varepsilon\bD_\Upsilon +\varepsilon'\bD_{\Upsilon+}' \pm \bD_\Upsilon^\times \\ 
    \bn'= \varepsilon'\bD_\Upsilon-\varepsilon\bD_{\Upsilon+}' ,\quad
    \dvec{t}' = \bd \bd_\Upsilon + \bd_\Upsilon \bd + \bd' \bd_\Upsilon' + \bd_\Upsilon' \bd'. 
\end{split}
\end{equation}
With these definitions, Eq.~\eqref{eq:p_nu'} then leads to 
the first Floquet correction $4\alpha^2 f_0 p_\nu'$ in Eq.~\eqref{eq:tildep2} to the undriven spin texture $\bar \bp$ 
\begin{equation}\label{eq:bp'}
\begin{split}
    \bp'&=\gamma_0(\varepsilon\bD_\Upsilon-\bd\times\bD_{\Upsilon-}') - \bd'\cdot\dvec{t}' - a_0^{-\prime}\bd' \\
    &+ \gamma_0^2 \bD_{\Upsilon+}' + \varepsilon \bD_{\Upsilon+}'\times\bd' + \delta \bp.
\end{split}
\end{equation}
It reads for the $\gamma_1$ case
\begin{equation}\label{eq:bp'_gamma1}
\begin{split}
    \bp'=
    \begin{cases}
    2\gamma_0 (0,\varepsilon d_2,-\gamma_1 d_2)+\bgamma_1  E_-  & \nu=1 \\
    2\gamma_0 (\varepsilon d_1,0,0)+\bgamma_1  (2\varepsilon^2-E_-)  & \nu=-1 \\
    \delta \bp(\tau) + 2\gamma_0 (\varepsilon d_1,\varepsilon d_2,-\gamma_1 d_2)+2\bgamma_1   \varepsilon^2  & \tau=\pm1
    \end{cases}
\end{split}
\end{equation}
with $E_\pm = \varepsilon^2-d^2 \mp(\gamma_0^2-\gamma^2)$. The effect of $\delta \bp(\tau=\pm1)=[\bar\bp(\delta m(\tau))-\bar\bp(m=0)]/4\alpha^2$, separated out and due to Eq.~\eqref{eq:delta_m0}, 
is highlighted in \ref{app:Flo_mass_effect} and thereafter absorbed in $\bar\bp$.

The second term in Eq.~\eqref{eq:G^<1} will directly give, with the help of Eq.~\eqref{eq:r<a}, 
\begin{equation}\label{eq:p^+gamma}
\begin{split}
    4p_0^+ &=\Gamma^+_0a_0^+ + \bGamma^+\cdot\ba^- \\
    4\bp^+ &= \Gamma^+_0\ba^+ + a_0^-\bGamma^+ + \bn\times\bGamma^+ + \bGamma^+\cdot\dvec{t}.
\end{split}
\end{equation}
For $\bp^-$, i.e., the contribution from the Floquet replica $n=-1$,
from Eq.~\eqref{eq:Gamma^pm}, we know that the index $s=\pm$ of $\Gamma^s$ effectively only reverses the polarization $\tau$ of the pumping light. This means that $\bp^-(\tau)=\bp^+(-\tau)$
are the same except that the expressions of $\tau=\pm1$ are exchanged. 
For the $\gamma_1$ case of our main interest, the spin texture expression is $4\bp^+\mycomment{(\gamma_1)}=A_1\bd+\bd\times\bB_1 - C_1\bgamma_1 + E_- \bGamma^+ + 4\varepsilon\gamma_0\gamma_1\tau\mycomment{\chi}\hat{y}$ 
where $A_1=2\Gamma^+_\nu\lambda^\nu,\,
    \bB_1= -4\gamma_0(\gamma_1,0,\gamma_0\tau\mycomment{\chi}),\,C_1=2[\gamma_0^2(1+\tau^2)+\gamma_1^2(1-\tau^2)]$, 
leading 
to Eq.~\eqref{eq:p^+_gamma1_main}.

\section{Generic properties of the system}
\subsection{Spin magnitude profile}\label{app:magnitude}

We analyze the spin texture magnitude below. Based on Eq.~\eqref{eq:G^<1}, the major variation in magnitude lies in the positive definite energy denominator Eq.~\eqref{eq:epsilon_denom0}.
Below, we discuss its profile Eq.~\eqref{eq:epsilon_denom_gamma1} for the $\gamma_1$ case.
Given a fixed energy $\varepsilon$-plane, the signal eventually decays at large momentum $k$ and the presence of $\gamma_1$ breaks 
the circular symmetry associated with the $\gamma_0$-only case. The profile satisfies the mirror symmetry with respect to $k_y$-axis, i.e., $\rho_0(k_x)=\rho_0(-k_x)$, which is also consistent with the general symmetry Eq.~\eqref{eq:Fsymm0}. However, in the orthogonal direction, the mirror symmetry with respect to $k_x$-axis, i.e., $\rho_0(k_y)=\rho_0(-k_y)$, only holds in the $\varepsilon=0$ energy plane. 

Hence, moving $\varepsilon$ can largely alter the magnitude profile. 
In fact, the presence of $\gamma_1$ in $E_{\gamma_1}^2$ is minor; the second square $(\varepsilon\gamma_0- v\hbar k_y \gamma_1)^2$ with $ \gamma_1 k_y$ makes a difference in the profile. The strength decays faster along $k_y$-axis than $k_x$-axis; more importantly, there is an $\varepsilon$-dependent movement of the strongest signal along the $k_y$-axis in order to keep the square small, which, as shown in Fig.~\ref{Fig:denominator_profile}, is roughly a \textit{crescent}-shaped blue region and still coincides with a part of the on-shell ring/annulus enhancement near the Dirac cone. We refer to this crescent region later, especially when discussing the spin texture in the $\mu_0\sim0$ case.

\subsection{Dichroic mass effect}\label{app:Flo_mass_effect}
Before going to the wider range of Floquet-induced spin textures, we discuss the manifestly dichroic effect of Eq.~\eqref{eq:delta_m0}, i.e., the dynamical mass generated by CP light pumping.
\begin{equation}
    \grave \bp=\bar\bp +4\alpha^2\delta \bp(\tau=\pm1)
\end{equation}
is effectively identical to the undriven texture 
when this effective mass is substituted since we are concerned about the massless case. According to \ref{itm:1}, as long as the occupation $f_0$ is finite, the undriven massless $\bar\bp$ dominates over other contributions in Eq.~\eqref{eq:tildep2} except for $\delta \bp$. Henceforth, without causing any confusion and for the sake of notational brevity, we will simply denote $\grave\bp$ by $\bar\bp$ with the correction understood.
Here, for completeness, we comment on both the $\gamma_1$ and $\gamma_3$ cases, respectively corresponding to the two representative undriven cases, i.e., the meron/vortex pair texture and the skyrmion-monopole texture.

\begin{itemize}

\item Dichroic meron-antimeron pair configuration: $\gamma_1$ relaxation

The mass generation will give in Eq.~\eqref{eq:bp'_gamma1}
\begin{equation}
        \delta p_3=2(\gamma_0\varepsilon-\mycomment{\chi}\gamma_1v\hbar k_y)\delta m'(\tau)
    \end{equation}
with $\delta m'(\tau)=\tau\Omega$.
From $p_3'\mycomment{=\delta p_3-\gamma_0(\bd\times\bD_{\Upsilon-}')_3} = \delta p_3-2\gamma_0\gamma_1 d_2$ and $p_3^+ = \gamma_0\gamma_1d_2-\frac{\tau\mycomment{\chi}}{2}\gamma_0 E_-$,
we see that all possible corrections to the undriven $\bar p_3$ are of the magnitude $\varepsilon^2\gamma_0$ or $\varepsilon\gamma_0\gamma_1$ to the leading order except for $\delta p_3$. As long as $\Omega\gg\gamma_1$ and we focus on the low-energy region, $\delta p_3$ dominates. This means that we effectively have the mass term purely generated and controlled by the CP light, including its sign and magnitude. For instance, CP light turns the vortex pair in the undriven massless case to a meron-antimeron pair and can switch the meron and antimeron in the pair by switching $\tau=\pm1$.

\item Dichroic skyrmion helicity: $\gamma_3$ relaxation 

The undriven $\gamma_3$ case has the spin texture $\bar\bp_{12}=2(\gamma_0\varepsilon+m\gamma_3)\bd_{12},
    \bar p_{3}=\gamma_3[\varepsilon(\varepsilon+2m\gamma_0/\gamma_3)+m^2-(\gamma_0^2-\gamma_3^2)-k^2]$. 
Similarly, we have the mass $m$ 
effectively given by the dichroic $\delta m(\tau)$. An interesting effect according to 
the skyrmion formation in this case is the dichroic skyrmion helicity given by 
\begin{equation}
    \mathcal{H}=\sgn[\gamma_0\varepsilon+\delta m\gamma_3\mycomment{(\gamma_0\varepsilon+\delta m\gamma_3)\chi}]\frac{\pi}{2}.
\end{equation}
In the $\varepsilon=0$-plane, $\mathcal{H}$ directly switches sign with the CP light $\tau=\pm1$. Even when $\varepsilon$ is finite, e.g., $\varepsilon<0$ in the photoemission-active region if $\mu_0\sim0$, 
$\mathcal{H}$ can switch its sign from $\frac{\pi}{2}$ to $-\frac{\pi}{2}$ at $\varepsilon=-\delta m\gamma_3/\gamma_0$ as $\varepsilon$ goes downward from 0 when $\tau=1$; on the other hand, $\mathcal{H}$ is fixed to $-\frac{\pi}{2}$ when $\tau=-1$.

\end{itemize}

\subsection{Symmetry relations}\label{app:symmetry}
Here we summarize the symmetry relations of the spin textures for the $\gamma_1$ case as observed in the full Floquet numerics. We have the general symmetry relation valid \textit{regardless} of the position of chemical potential $\mu_0$
\begin{equation}\label{eq:Fsymm0}
\begin{split}
    p_{0,1}\left(-\tau ,\varepsilon ,-k_x,k_y\right)&=p_{0,1}(\tau ,\varepsilon ,\bk)\\
    -p_{2,3}\left(-\tau ,\varepsilon ,-k_x,k_y\right)&=p_{2,3}(\tau ,\varepsilon ,\bk).
\end{split}
\end{equation}
For instance, in the case of the pure pumping-induced texture  
and the $\mu_0\sim0$ in the main text, 
this is the only relation available. 
For $\mu_0>\Omega$ in the main text, 
we additionally have two independent symmetry properties
\begin{equation}\label{eq:Fsymm1}
\begin{split}
p_{0,1,2}(-\tau ,\varepsilon ,\bk)&=p_{0,1,2}(\tau ,\varepsilon ,\bk)\\
-p_3(-\tau ,\varepsilon ,\bk)&=p_3(\tau ,\varepsilon ,\bk)
\end{split}
\end{equation}
and 
\begin{equation}\label{eq:Fsymm2}
    p_{\mu }(-\tau ,-\varepsilon ,-\bk)=p_{\mu }(\tau ,\varepsilon ,\bk).
\end{equation}
Note that $p_{\mu }\left(\pm \tau ,\varepsilon ,-\bk\right)\neq p_{\mu }\left(\tau ,\varepsilon ,\bk\right)$ in general and all these relations as well hold in the special $\varepsilon=0$-plane and the LP light case ($\tau=0$).

Now we can use the analytical expressions in the main text 
to justify these symmetry properties. 
When the chemical potential is high as in the $\mu_0>\Omega$ case, 
the total spin texture becomes
\begin{widetext}
\begin{equation}\label{eq:tildep_gamma1}
\begin{split}
    \bp&= \bar\bp+4\alpha^2(\bp'+\bp^++\bp^-)=\bar\bp +4\alpha^2 \begin{cases}
    (\gamma_0\varepsilon-\gamma_1d_1)\bd_{12}+(\frac{1}{2}E_+-2\gamma_1^2)\bgamma_1+2\gamma_0\varepsilon \bd_2 & \nu=1 \\
    (\gamma_0\varepsilon+\gamma_1d_1)\bd_{12}+(\frac{3}{2}E_++2d^2)\bgamma_1+2\gamma_0\varepsilon \bd_1 & \nu=-1 \\
    \delta \bp(\tau)+4\gamma_0\varepsilon \bd_{12}+2\bgamma_1(\varepsilon^2-\gamma_0^2) & \tau=\pm1
    \end{cases}.
\end{split}
\end{equation}
\end{widetext}
It is straightforward to see that Eq.~\eqref{eq:tildep_gamma1} satisfies all the three independent symmetry relations Eqs.~\eqref{eq:Fsymm0}\eqref{eq:Fsymm1}\eqref{eq:Fsymm2}. Besides, each of $\bp',\bp^\pm,\bar\bp$, i.e., Eqs.~\eqref{eq:bp'_gamma1}\eqref{eq:p^+_gamma1_main} and also the undriven spin texture, satisfies the most general symmetry relation Eq.~\eqref{eq:Fsymm0}. This explains, to the extent of low-energy theory, why Eq.~\eqref{eq:Fsymm0} holds independently of $\mu_0$ while Eqs.~\eqref{eq:Fsymm1}\eqref{eq:Fsymm2} hold only for the case when $\mu_0>\Omega$. 

Numerically, we also observe the following. i) LP light does not induce $S_z$ when $\mu_0>\Omega$, i.e., $p_3(\tau =0)\equiv0$, which is implied also by Eq.~\eqref{eq:Fsymm1}; ii) however, this is not the case for generic lower $\mu_0$ and $4\alpha^2\Omega\sim|p_3(\tau =0,\nu=1)|\gg |p_3(\tau =0,\nu=-1)|\neq 0$. Firstly, i) is consistent with Eq.~\eqref{eq:tildep_gamma1} with $\delta m=0$.
Secondly, for ii) when $\mu_0$ is not high, the contribution from the replicas of upper and lower dispersion branches are not necessarily fully cancelled, which is obvious in the finite $p_3',p_3^+$ for $\nu=1$ in Eqs.~\eqref{eq:bp'_gamma1}\eqref{eq:p^+_gamma1_main}. Thirdly, although they can vanish for $\nu=-1$ in the low-energy theory, it holds only up to the leading order within the perturbation theory. Note also that ii) does not contradict the appearance of Bloch lines in the main text, 
which only means that $p_3(\tau=0)$ can vanish at special momentum points dependent on the energy plane.

\section{Analysis of spin textures}
\subsection{Pure pumping texture}\label{app:p^+}
We show how the pure pumping texture $\bp^+$, 
given in Eq.~\eqref{eq:p^+_gamma1_main}, is analyzed.

We first calculate below to find two cases when $\bp^+_{12}=0$ for the XLP case.
\begin{itemize}
\item $v\hbar k_y=-\frac{\gamma_0}{\gamma_1}\varepsilon$ 

We have $p^+_2=0$ and
\begin{equation}
\begin{split}
    p^+_1=\frac{1}{4\gamma_1}[(\gamma_0^2-\gamma_1^2)(\varepsilon^2-\varepsilon_0^2)+v^2\hbar^2\gamma_1^2k_x^2]
\end{split}
\end{equation}
with $\varepsilon_0^2=\frac{\gamma_1^2}{\gamma_0^2-\gamma_1^2}(3\gamma_0^2+\gamma_1^2)$.
Then, $p^+_1=0$ has two solutions
\begin{equation}
    v\hbar k_x(\varepsilon)=\pm\Delta_-(\varepsilon)/\gamma_1
\end{equation}
and hence $p^+_3\neq0$ in any $\varepsilon$-plane within the range
\begin{equation}\label{eq:range_vortex}
    -\varepsilon_0<\varepsilon<\varepsilon_0,
\end{equation}
whereby we denote the square root of the discriminant also for later use
\begin{equation}
    \Delta_s(\varepsilon)=\sqrt{s(\gamma_0^2-\gamma_1^2)(\varepsilon^2-\varepsilon_0^2)}.
\end{equation}
For the coordinate $\bk'$ measured from the cores $\bK^\pm(\varepsilon)=(k_x=\pm\Delta_-(\varepsilon)/v\hbar \gamma_1,k_y=-\frac{\gamma_0}{\gamma_1}\varepsilon)/v\hbar $, the spin texture takes the form
\begin{equation}
\begin{split}
    \bp^+_\pm(\bk')&=\frac{1}{4}\left(  \pm2\Delta_-(\varepsilon)v\hbar k_x'+v^2\hbar^2 \gamma_1(k_x'^2-k_y'^2),\right.\\
    &\left.2(\pm\Delta_-(\varepsilon)+v\hbar \gamma_1k_x')k_y', 4\gamma_0(\pm\Delta_-(\varepsilon)+v\hbar \gamma_1k_x')\right).
\end{split}
\end{equation}
This, to the leading order, readily gives a meron-antimeron pair
\begin{equation}\label{eq:p^+meron}
    \tilde\bp^+_\pm(\bk')=\pm\frac{1}{2}\Delta_-(\varepsilon)\left( v\hbar k_x', v\hbar k_y', 2\gamma_0 \right).
\end{equation}
Therefore, in the 3D $(\bk,\varepsilon)$-space and within $-\varepsilon_0<\varepsilon<\varepsilon_0$, a pair of meron-antimeron strings along $(\varepsilon,\pm\Delta_-(\varepsilon)/\gamma_1,-\frac{\gamma_0}{\gamma_1}\varepsilon)$ merges at $\bX^\pm=(\pm\varepsilon_0,0,\mp\frac{\gamma_0}{\gamma_1}\varepsilon_0)$.

\item $k_x=0$

We have $p^+_2=p^+_3=0$ in the first place. Also, the first component
\begin{equation}
    p^+_1=-\frac{1}{4}[(v\hbar)^2 \gamma_1k_y^2+2v\hbar \gamma_0\varepsilon k_y+\gamma_1(\varepsilon^2+3\gamma_0^2+\gamma_1^2)]
\end{equation}
can vanish in an $\varepsilon$-plane at two solutions
\begin{equation}
    v\hbar k_y(\varepsilon)=\frac{-\gamma_0\varepsilon\pm\Delta_+}{\gamma_1}
\end{equation}
as long as its discriminant
\begin{equation}
    \Delta_+^2=(\gamma_0^2-\gamma_1^2)(\varepsilon^2-\varepsilon_0^2)\geq0,
\end{equation}
which is exactly the complement of Eq.~\eqref{eq:range_vortex}. 
For the coordinate $\bk'$ measured from the cores $\bK^\pm(\varepsilon)=(0,k_y=(-\gamma_0\varepsilon\pm\Delta_+)/v\hbar \gamma_1)$, the spin texture takes the form
\begin{equation}
\begin{split}
    \bp^+_\pm(\bk')&=\frac{1}{4}\left(  \mp2\Delta_+(\varepsilon)v\hbar k_y'+(v\hbar )^2\gamma_1(k_x'^2-k_y'^2),\right.\\
    &\left.2v\hbar (\pm\Delta_+(\varepsilon)+v\hbar \gamma_1k_y')k_x', 4v\hbar \gamma_0\gamma_1k_x'\right).
\end{split}
\end{equation}
This, to the leading order, readily gives an in-plane vortex pair
\begin{equation}\label{eq:p^+BL}
    \tilde\bp^+_\pm(\bk')=\frac{1}{2}v\hbar \left(  \mp\Delta_+(\varepsilon)k_y', \pm\Delta_+(\varepsilon)k_x', 2\gamma_0\gamma_1k_x'\right).
\end{equation}
Therefore, we have a pair of Bloch lines along $(\varepsilon,0, (-\gamma_0\varepsilon\pm\Delta_+)/\gamma_1)$ also merged at the foregoing $\bX^\pm$.

\end{itemize}
There is a pair of special energy planes, $\varepsilon=\pm\varepsilon_0$. We briefly look at the behavior in the vicinity of $\bX^\pm$, i.e., where the foregoing meron strings and Bloch lines merge. For the coordinate $\bX'=(\varepsilon', k_x', k_y')$ measured from $\bX^\pm$, we have
\begin{equation}
\begin{split}
    \bp^+_\pm(\bX')&=\frac{1}{2}\left(
     \mp \tilde{\gamma}_1^2\varepsilon' + \frac{\gamma_1}{2} ((v\hbar )^2(k_x'^2- k_y'^2)-\varepsilon'^2) \right. \\
      &\left. - \gamma_0\varepsilon' v\hbar k_y',v\hbar k_x'(\gamma_0\varepsilon'+\gamma_1 v\hbar k_y'),\,
     2\gamma_0\gamma_1 v\hbar k_x'
    \vphantom{\frac{\gamma_1}{2}}\right)
\end{split}
\end{equation}
with $\tilde{\gamma}_1^2=\sqrt{(\gamma_0^2-\gamma_1^2)(3\gamma_0^2+\gamma_1^2)}$.
Around the anisotropic Bloch point $\bX^\pm$, it is linear to the leading order only along  $\varepsilon'$ for $p^+_1$ and along $k_x'$ for $p^+_3$.

It is clear that the above features owe fully to the finite NH $\gamma_1$-relaxation. In fact, in the limit of $\gamma_1=0$, we have $\varepsilon_0\rightarrow0$ and the meron ring in Fig.~\ref{Fig:big_pure_tau=0}(a) shrinks towards the origin $\bX=0$ and disappears; the Bloch lines are deformed to the union of $\varepsilon$-axis and $k_y$-axis, which merges into the union of  $\varepsilon$-axis and $\varepsilon=0$-plane where the signal $\bp^+(\bgamma=0)=\frac{1}{2}\gamma_0\varepsilon\bd$ vanishes.

We secondly inspect the CP light driving case by rewriting the spin texture in Eq.~\eqref{eq:p^+_gamma1_main} as 
\begin{equation}\label{eq:p^+_gamma1_tau=pm1}
\begin{split}
    \bp^+(\bk')&=
    \gamma_0v\hbar \left(\vphantom{\frac{\tau}{2}}-(\varepsilon k_y'+\gamma_0\tau k_x'),\varepsilon k_x'-\gamma_0\tau k_y',\right.\\
    &\left.-\frac{\tau}{2v\hbar }(\varepsilon^2-(v\hbar )^2k'^2+\gamma_0^2)\right),
\end{split}
\end{equation}
which is achieved by shifting the momentum to  $\bk'=\bk-\bK$ with
\begin{equation}\label{eq:p^+_CPL_Kinfo}
    \bK=(-\tau\gamma_1,0)/v\hbar 
\end{equation}
independent of $\varepsilon$-plane.
We then transform to the polar coordinate of $\bk'=k'(\cos\phi,\sin\phi)$ and can cast the in-plane spin as
\begin{equation}\label{eq:p^+_12}
\begin{split}
    \bp^{+}_{12}(\bk') &= v\hbar k'\gamma_0(-\gamma_0\tau\cos\phi-\varepsilon\sin\phi,\varepsilon\cos\phi-\gamma_0\tau\sin\phi) \\
    &= v\hbar k'\gamma_0\sqrt{\varepsilon^2+\gamma_0^2}\left(\cos(\phi+\phi_0),\sin(\phi+\phi_0)\right)
\end{split}
\end{equation}
with $\phi_0=\arctan{(-\gamma_0\tau,\varepsilon)}$, where the 2-argument $\arctan{(x,y)}$ gives the polar angle of a point $(x,y)$ on the two-dimensional plane. 
Hence, $\bp^{+}_{12}(\bk)$ is a vortex shifted by $\bK$ with vorticity $\mathcal{V}=1$ and helicity $\mathcal{H}=\phi_0$.
On the other hand, the third component 
\begin{equation}
    p^+_3(\bK) = -\frac{\tau\gamma_0}{2}(\varepsilon^2+\gamma_0^2) \lessgtr 0 \qquad (\tau=\pm1)
\end{equation}
never vanishes and has a fixed sign.
At point $(\varepsilon,\bK)$ in any $\varepsilon$-plane the spin points either upward ($\tau=-1$) or downward ($\tau=1$) and the $k'^2$-term will always reverse the sign of $p^+_3$ and also dominates over $\bp^+_{12}$ at large $k$. These properties constitute a skyrmion string.

\begin{figure*}[hbt]
\includegraphics[width=17.8cm]{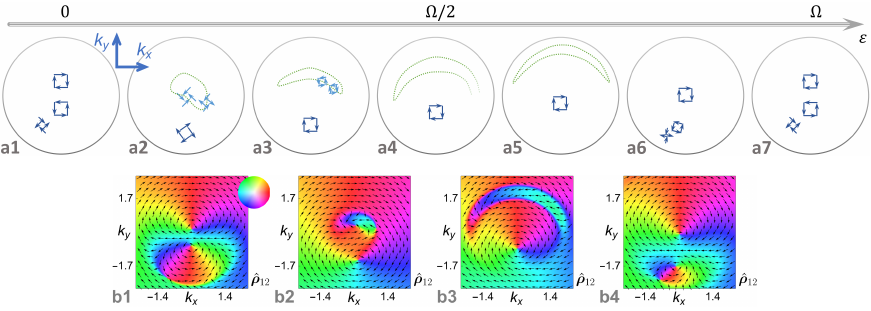}
\caption{\textbf{Energy-evolution of CP light-driven textures for low-energy chemical potential $\mu_0$.} In-plane momentum-space spin texture (a) schematics and (b) numerical data highlight the energy-plane evolution of vortices and domains for RCP light; LCP light is readily related by symmetry relations. 
Selected energy planes (b1-4) $\varepsilon/\Omega=-0.035,0.17,0.5,0.87$ roughly correspond to (a1,a2,a4,a6), respectively. Other parameters and presentation styles follow Fig.~\ref{Fig:big_mu0_tau=0}.
}\label{Fig:big_mu0_tau=1}
\end{figure*}

\subsection{\texorpdfstring{$\mu_0\sim0$}{Lmu} case}\label{app:mu=0}
Firstly, we show how the background texture as per \ref{itm:2} is found from the equilibrium $\bar\bp(\varepsilon-\Omega)$.
From its $\varepsilon$-dependent movement property and according to the vortex\mycomment{/meron} cores at 
\begin{equation}\label{eq:K^pm}
    \bK^\pm=\left(0,\frac{1}{v\hbar \gamma_1}\left[\mycomment{\chi}\gamma_0\varepsilon\pm\sqrt{(\gamma_0^2-\gamma_1^2)(\varepsilon^2+\gamma_1^2)\mycomment{+\gamma_1^2m^2}}\right]\right),
\end{equation}
one vortex core relevant here is shifted along the $-\hat{k}_y$-direction till $K_y^+(\varepsilon-\Omega)$, which satisfies $K_y^+(\varepsilon-\Omega)<0$ as long as $\varepsilon-\Omega<-\sqrt{\gamma_0^2-\gamma_1^2}$; the other vortex is much further away in the -$\hat k_y$-direction and not within the momentum observation window. Therefore, for the typical $\Omega\gg\sqrt{\gamma_0^2-\gamma_1^2}$ that makes $K_y^+(\varepsilon-\Omega)\ll0$, the large region $k_y>K_y^+(\varepsilon-\Omega)$ in our observation window falls in the $+\hat{k}_y$-direction with respect to the vortex core at $\bK^+$ in the undriven texture. To see this, for $0<\varepsilon\ll\Omega/2$ our low-energy observation window $v\hbar k\ll\Omega$ often falls well within the corresponding $\varepsilon$-dependent on-shell ring and we can take the limit  $|\varepsilon-\Omega|\gg v\hbar k$ of the undriven Eq.~\eqref{eq:p_nu_gamma1_maintext}
\begin{equation}\label{eq:bp_gamma1_inshell}
    \bar\bp(\varepsilon-\Omega)\approx2\gamma_0(\varepsilon-\Omega)\bd+(\varepsilon-\Omega)^2\bgamma_1.
\end{equation}
This is an in-plane vortex considerably shifted by $\gamma_1\frac{\varepsilon-\Omega}{2v\hbar \gamma_0}\hat{k}_y$ as expected. Then the background texture is in gross $-2\gamma_0k(-\sin\phi,\cos\phi,0)$ with the $\bk$-space polar angle $\phi\sim\pi/2$, i.e., spins pointing around $+\hat{x}$-direction.

Secondly, for \ref{itm:1}, at large-$(\bk,\varepsilon)$ region close to the on-shell region, 
$\bp^+$ takes on form below for the LP and CP cases
\begin{equation}\label{eq:p^+_gamma1_on}
\begin{split}
    \bp^+_\mathrm{on}=\begin{cases}
    \frac{1}{2}(\gamma_0\varepsilon-\nu\gamma_1d_1)\bd_{12} & \nu=\pm1 \\
    \gamma_0\varepsilon(d_1,d_2+\tau\gamma_1,0) & \tau=\pm1
    \end{cases}
\end{split}
\end{equation}
We explain the distinction between two orthogonal LP cases of Eq.~\eqref{eq:p^+_gamma1_on}.
The $k_y$-dependent prefactor due to finite $\gamma_1$ brings about a visible strength anisotropy when $\tau=0$, which is controlled by the type of LP light ($\nu=\pm1$). Crucially, in the $\nu=1$ ($\nu=-1$) case with $\varepsilon>0$, the on-shell crescent region [see Fig.~\ref{Fig:denominator_profile}(d,e) and \ref{app:magnitude}] of the largest signal magnitude is located around the point furthest away from (closest to) the $k_y=K_y^0$-line where $\bp^+_\mathrm{on}$ vanishes. 
This is entirely caused by the $\varepsilon$-dependent movement with the moving direction controlled by $\nu$. Hence, we typically have $|\bp^+_\mathrm{on}(\nu=1)|\gg|\bp^+_\mathrm{on}(\nu=-1)|$ for the crescent region.

Thirdly, we detail how Eq.~\eqref{eq:bp_mu0_main} captures the light blue vortex pair in Fig.~\ref{Fig:big_mu0_tau=0}(a2-4). The corresponding analytical expression combines Eq.~\eqref{eq:p^+_gamma1_main} and the background $\bar\bp(\varepsilon-\Omega)$
\begin{equation}\label{eq:bp_mu0}
\begin{split}
    \bp(\bk,\varepsilon) &= \frac{4\alpha^2\bp^+(\varepsilon,\nu=1)}{W(\varepsilon)} + \frac{\bar\bp(\varepsilon-\Omega)}{W(\varepsilon-\Omega)} \\
    &= \frac{\frac{1}{2}(\gamma_0\varepsilon-\gamma_1d_1)\bd_{12}-(\frac{1}{4}E_++\gamma_0^2)\bgamma_1+\gamma_0\gamma_1d_2\hat{z}}{(4\alpha^2)^{-1}\,W(\varepsilon)} \\
    &+ \frac{2(\gamma_0(\varepsilon-\Omega)+\gamma_1d_1)\bd_{12}+E_+(\varepsilon-\Omega)\bgamma_1}{W(\varepsilon-\Omega)}.
\end{split}
\end{equation}
We readily notice that $p_2=0$ as long as $d_2=v\hbar k_x=0$. Thus, the vortex cores, which is a Bloch point, are the points where $p_1(d_1=-v\hbar k_y)=0$ is also satisfied. This quadratic equation exactly bears two solutions: they correspond to the light blue pair of vortices along $k_y$-axis and mark the edges along the $k_y$-axis of the domain region in Fig.~\ref{Fig:big_mu0_tau=0}(a2-4) and the corresponding Fig.~\ref{Fig:big_mu0_tau=0}(b2,b3).

\subsubsection{CP light pumping: dichroically tilted open domains}

Now we discuss the CP light pumping case that shows clear dichroism. 
The second line of Eq.~\eqref{eq:p^+_gamma1_on} acquires a $\tau$-dependent vortex core position $\bK(\tau)=(-\tau\gamma_1,0)/v\hbar $. 
This leads to dichroic tilting of the crescent region, i.e., the crescents in Fig.~\ref{Fig:denominator_profile}(d,e) will be slightly rotated clockwisely (counterclockwisely) for $\tau=\pm1$ [see also Fig.~\ref{Fig:big_mu0_tau=1}(a2,a3)], simply because the core where in-plane spin vanishes is slightly shifted to the $\mp\hat{k}_x$-direction. Beyond this on-shell approximation, the $\tau$-dependent helicity $\mathcal{H}(\tau)$ of $\bp^+$ also adds to such asymmetry [c.f. Eq.~\eqref{eq:p^+_12}]. 
Such tilted crescents naturally induce dichroic domains, where the in-plane major spin pointing approximately matches the predicted helicity difference $\mathcal{H}(\tau=1)-\mathcal{H}(\tau=-1)=\pi/2$.
Interestingly, because of the tilting, 'open domain' can occur at high energies $\varepsilon\sim\Omega/2$ in Fig.~\ref{Fig:big_mu0_tau=1}(a2-4) and corresponding Fig.~\ref{Fig:big_mu0_tau=1}(b2,b3). This is caused by the pair annihilation of light blue up/down in-plane vortices, corresponding to the $\nu=1$ counterpart now rotated and distorted to the $\pm\hat{k}_x$-direction.

\subsection{\texorpdfstring{$\mu_0>\Omega$}{Hmu} case}\label{app:mu>Omega}
Here, we show how the vortex configurations are found analytically. 
Firstly, we inspect the $\varepsilon=0$-plane. 
With the approximation $\bar\bp(0) \approx \frac{2\gamma_1d_1\bd-d_+^2\bgamma_1}{d_+^4}$ and $
    \bar\bp(\mp\Omega) \approx \frac{2(\mp\gamma_0\Omega+\gamma_1d_1)\bd-(\Omega^2-d_+^2)\bgamma_1}{(\Omega^2-d_+^2)^2}$ valid when $\Omega\gg\gamma_{0,1}$, we find
\begin{equation}\label{eq:p(0)}
\begin{split}
    \bp(0) \approx \frac{[(\Omega^2-d_+^2)^2+2d_+^4]2\gamma_1d_1\bd}{d_+^4(\Omega^2-d_+^2)^2} + \frac{(3d_+^2-\Omega^2)\bgamma_1}{d_+^2(\Omega^2-d_+^2)} 
\end{split}
\end{equation}
with $d_+^2=d^2+(\gamma_0^2-\gamma_1^2)$.
Two in-plane vortex solutions $\bp_{12}(\varepsilon=0,\bK^\pm)=0$ located at $\bK^\pm=(\pm[\Omega^2/3-(\gamma_0^2-\gamma_1^2)]^\frac{1}{2},0)/v\hbar $ follow.
To see \eqref{eq:muH_vortexinfo1} from Eq.~\eqref{eq:p(0)}, one can readily check the azimuthal angle $\Theta(\bp)$ of the spin texture in four directions $\phi=0,\pi/2,\pi,3\pi/2$ around the vortex core with $\phi$ the polar angle in $\bk$-space measured in the vicinity of $\bK^\pm$. 
Corresponding to Fig.~\ref{Fig:big_muH_tau=0}(a1,b1), we find that $\Theta^\pm(\varepsilon=0,\phi=0)=0,\pi,
\Theta^\pm(\varepsilon=0,\phi=\pi)=\pi,0$ and $\Theta^\pm(\varepsilon=0,\phi=\pi/2)=3\pi/2,\pi/2, \Theta^\pm(\varepsilon=0,\phi=3\pi/2)=\pi/2,3\pi/2$.

The $\varepsilon=\Omega/2$-plane is analyzed similarly to Eq.~\eqref{eq:p(0)}. In this case, we can drop in Eq.~\eqref{eq:bp_Hmu} $\bar\bp(\varepsilon+\Omega)$ from Floquet replica $n=-1$ since $\varepsilon=\Omega/2$-plane lies right at the middle between the $n=0,1$ replicas in Fig.~\ref{Fig:Floquet} and other replicas' contribution is suppressed by their large energy denominator $W$. 
We thus find the following expression to the leading order in small $|k_y|$
\begin{equation}\label{eq:p(Omega/2)}
\begin{split}
    &\bp(\Omega/2) 
    = \,\bar\bp(\Omega/2)+\bar\bp(-\Omega/2) \\
    \approx \,&\frac{[(\frac{\Omega^2}{4}-d_+^2)^2-\gamma_0^2\Omega^2]4\gamma_1d_1\bd}{[(\frac{\Omega^2}{4}-d_+^2)^2+\gamma_0^2\Omega^2]^2} + \frac{2(\frac{\Omega^2}{4}-d_+^2)\bgamma_1}{(\frac{\Omega^2}{4}-d_+^2)^2+\gamma_0^2\Omega^2}.
\end{split}
\end{equation}
Two in-plane vortex solutions $\bp_{12}(\varepsilon=\Omega/2,\bK^\pm)=0$ with $\bK^\pm=(\pm[\Omega^2/4-(\gamma_0^2-\gamma_1^2)]^\frac{1}{2},0)/v\hbar $ lead to Eq.~\eqref{eq:muH_vortexinfo2}.
To see this, we again check the azimuthal angle $\Theta(\bp)$ and find $\Theta^\pm(\varepsilon=\Omega/2,\phi=0)=\pi,0, \Theta^\pm(\varepsilon=\Omega/2,\phi=\pi)=0,\pi$ and $\Theta^\pm(\varepsilon=\Omega/2,\phi=\pi/2)=\pi/2,3\pi/2,
\Theta^\pm(\varepsilon=\Omega/2,\phi=3\pi/2)=3\pi/2,\pi/2$.

\end{document}